\documentclass[11pt]{article}
\usepackage[utf8]{inputenc}
\pdfoutput=1
\usepackage{amssymb,amsmath,mathrsfs,enumerate}
\usepackage{graphicx,rotate,multicol}
\usepackage{float}
\usepackage{tocloft}
\usepackage{subfig}
\usepackage[margin=10pt,labelfont=bf]{caption}
\usepackage{cite}
\usepackage{soul}
\usepackage[colorlinks=true,
            linkcolor=blue,
            urlcolor=blue,
            citecolor=teal]{hyperref}
\usepackage{multirow}
\usepackage{placeins}

\makeatletter
\let\Hy@linktoc\Hy@linktoc@page
\makeatother
\usepackage{color}
\definecolor{ourcolor}{rgb}{0.7, 0.25, 0.05}
\usepackage{tikz,braket}
\long\def\rpl#1!!#2!!{\textcolor{red}{#1} \textcolor{blue}{#2}}

\def \order(#1){{\mathcal O} \left(#1 \right)}

\evensidemargin=\oddsidemargin
\allowdisplaybreaks

\title{\color{black}{\bf Constraints on dark matter self-interaction from galactic core size}}

\author {\bf Tirtha Sankar Ray,$^{a,}$\footnote{tirthasankar.ray@gmail.com} 
	\hspace{4pt}  Sambo Sarkar$^{a,}$\footnote{sambosarkar92@gmail.com} \hspace{3pt} and Abinash Kumar Shaw$^{a,b,}$\footnote{abinashkumarshaw@gmail.com}
	\\[10pt]
	\small\em $^a$Department of Physics, Indian Institute of Technology Kharagpur, Kharagpur 721302, India \\
\small\em $^b$Astrophysics Research Centre, Open University of Israel, Ra'anana 4353701, Israel.	
}

\date{}

\begin{document} 
	
	\maketitle

	\begin{abstract}
		
		Self-interaction of particulate dark matter may help thermalising the central region of the galactic halo and driving core formation. The core radius is expectedly sensitive to the self-interaction strength of dark matter (DM). In this paper we study the feasibility of constraining dark matter self-interaction from the distribution of the core radius in isolated haloes. We perform systematic DM only $N$-body simulations of spherically symmetric isolated galactic haloes in the mass range of $10^{10} $-$10^{15}M_{\odot}$, incorporating the impact of isotropic DM self-interaction. Comparing the simulated profiles with the observational data, we provide a conservative upper limit on the self-interaction cross-section, $ \sigma/m < $ $ 9.8 $ $\ \rm cm^2 /\rm gm $ at $ 95 \% $ confidence level. We report significant dependence of the derived bounds on the galactic density distribution models assumed for the analysis.

\end{abstract}

\section{Introduction}
\label{sec:intro}

The standard cosmological framework Lambda Cold Dark Matter ($\Lambda$CDM) \cite{1984Natur.311..517B} has proven to be extremely successful in explaining the formation and evolution of large scale structures in our Universe \cite{Lisanti:2016jxe, Pace:2019vrs, DelPopolo:2002sz}. However, at the galactic scales there exists several mismatch between simulations and observations \cite{Bernardeau:2001qr}. The core-cusp \cite{BoylanKolchin:2003sf} and missing satellite problem \cite{BoylanKolchin:2011de} are some of the concerns that have been discussed in this context. Admittedly, these discrepancies may be a reflection of various limitations in the present day $N$-body simulations. It has been extensively discussed that a more careful consideration of baryonic effects may ease most of the tensions \cite{10.1111/j.1365-2966.2012.20571.x,Kim:2017iwr,10.1111/j.1365-2966.2010.16613.x}. However, these may also be indicative of specific microscopic properties of dark matter (DM) itself. In this paper we adopt this latter paradigm and investigate the correlation of observed galactic core and strength of self-interaction of particulate DM.

Collision-less cold DM (CDM) in the  absence of baryonic physics is expected to get gravitationally cooled in the neighborhood of the galactic center  during galaxy formation, aiding further collapse and eventually leading to a cusp like structures \cite{Sean:2017sea}. However, astrophysical observations \cite{Kamada:2016euw} spanning from dwarf spheroidal  galaxies to galaxy clusters indicate that DM haloes are less dense in their central regions as compared to predictions from collision-less CDM $N$-body simulations \cite{Flores:1994gz,Burkert:1995yz}. A viable possibility to address  this problem is to consider self-interaction within the dark sector \cite{Spergel:1999mh} that are widely motivated by particle physics models \cite{Khlopov:2013ava,Huo:2019yhk, Kamada:2020buc}. These self-interactions can thermalise the collapsing galactic center thus resisting formation of cusps \cite{Rocha:2012jg}. In this paper we focus on constraining the DM self-interaction cross-section by comparing the results from DM only $N$-body simulation of spherically symmetric isolated cores with observations in a wide mass range of astrophysical objects, ranging from galaxies  to clusters. This is complementary to the several studies that constrain self-interacting dark matter (SIDM) models using observational evidence  from rotation curves \cite{Elbert:2014bma,10.1093/mnrasl/sls053,Bondarenko:2017rfu}, stellar kinematics \cite{Correa_2021} and gravitational lensing \cite{Randall:2007ph,Peter:2012jh,Harvey:2015hha,Banerjee:2019bjp} alongside $N$-body simulations.

The paper is organized as follows. In section \ref{sec:ccSIDM} we review the rationale for SIDM as a possible solution to the core-cusp problem before we discuss the $N$-body simulations performed using the SIDM framework for isolated haloes. In section \ref{sec:OBScore} we discuss the observational data which is used to compare the results of our $N$-body simulations. In section \ref{sec:results} we present the obtained bounds on the self-interaction cross-section of DM from galactic core radius before concluding in section \ref{sec:conc}.

\section{The SIDM  solution to the core cusp problem}
\label{sec:ccSIDM}

$N$-body simulations, based on the collision-less CDM framework usually lead to an over-dense central region inside the DM haloes. Gravitational collapse of galaxies into the central potential of the halo causes thermal cooling in the core region. This reduces the thermal pressure that can resist the collapse, eventually leading to a \textit{cusp} like galactic center. However, observations reveal a somewhat different and more nuanced  picture. The rotation curves of low surface brightness galaxies and late-type disk like, gas-rich dwarf galaxies that are majorly DM dominated  \cite{2010AdAst2010E...5D}, indicate the presence of a constant-density or mild cusp-like galactic center with a logarithmic slope of $\sim 0.2$. Cored DM profiles have also been inferred for the more luminous spiral galaxies \cite{Salucci:2007tm}.  Previous works like \cite{Nakano:1997sn} have also advocated the presence of weak density cusps with a logarithmic slope $\sim 0.5 $ in large elliptical galaxies, motivated from the results of HST observations. Isolating the DM from the stellar components indicate an inner logarithmic density slope $\sim 0.5 \pm 0.13$ for clusters having weak cores \cite{Newman_2013n}. Alternately many recent works in literature tend to resolve the dichotomy of DM density inside the galactic center by rigorous stellar modeling \cite{Tollet_2016}, studying the dynamics of dwarf galaxies \cite{Read:2018pft}, using stellar kinematics and HI rotation curves \cite{Read:2018fxs}, cosmological hydrodynamic simulations \cite{Oman:2017vkl} etc. The core-cusp problem is the accepted moniker for this discrepancy in literature \cite{Flores:1994gz}.

Self-interaction of DM may be instrumental in thermalising the core by aiding  transport of heat  from the outer edges of the galaxies to the dense central region. This thermalisation of the central region of the DM halo causes the matter to get uniformly distributed within the region \cite{Kamada:2019wjo}. As we move away from the core the density falls off, reducing the self-interaction rate and ultimately behaves like an effective collision less CDM towards the less dense outer edges of the isolated haloes. A typical estimate for the core size may be obtained by requiring at least one self-interaction event per particle within the core, during the lifetime history of the galactic evolution \cite{Kaplinghat:2015aga},
\begin{equation*}
\label{eq:a}
\frac{\langle \sigma v \rangle_{self}}{m}\, \rho(r_c)\,  t_{\rm age} \approx 1 \, , 
\end{equation*}
where $\langle \sigma v \rangle_{self}$ denotes the velocity weighted self-interaction cross section, $m$ is the DM particle mass, $\rho(r)$ is the DM density profile,  $t_{\rm age}$ is the time for evolution of the virialized object and $r_c$ is the estimated core radius.  As evident from this estimate, the core size being a quantifying parameter of the galactic halo, can be directly related to the self-interaction strength of the DM. The possibility of constraining DM self-interaction from the core radius has been previously presented in \cite{Sokolenko:2018noz}. It has been pointed out that the surface density of the DM haloes \cite{Andrade:2020lqq} and  the velocity dispersion in the central regions can be used to constrain the DM self-interaction \cite{Peter:2012jh,Sagunski:2020spe}.

Thus the observed distribution of core radius in various isolated haloes can translate to a limit on the DM self-interaction.  As a proof of principle in this paper we perform $N$-body simulations of isolated haloes using a modified  \texttt{GADGET} \cite{Springel:2000yr,Springel:2005mi,Jun:2011jun} simulation tool, taking into account the leading effect of DM self-interaction. We consider a simplified phenomenological SIDM framework where velocity independent elastic $2\to 2$ DM scatter process dominate the DM self-interactions.  By comparing the distribution of core radius as function of the galaxy mass in simulated results and observed data, we determine a conservative upper-limit on the DM self-interactions. 

\subsection{N-body simulation of isolated SIDM haloes}
\label{sec:Nbodyint}

A few comments about the $N$-body simulations carried out in this work is now in order. To simulate the isolated haloes by incorporating the effects of the DM self-interaction, we utilize a modification of the existing gravitational $N$-body code \texttt{GADGET} \cite{Springel:2000yr,Springel:2005mi}. The algorithm follows a basic assumption that whenever a particle (say $j$) is considered to be the nearest neighbour of another particle (say $i$), there exists a finite probability of scattering or interaction. If and when the two particles collide in this sense, the phase space of the  scattered DM particles is   obtained from the differential scattering cross section $ \frac{\rm d \sigma}{\rm d \Omega}$. The positions and velocities are updated at each time-step following the usual Kick-Drift-Kick Leap-frog method.

We generate the initial phase-space of DM particles with a NFW matter distribution. For this purpose we use the publicly available code \texttt{HALOGEN} \cite{Zemp:2007nt} which sets up spherically symmetric distributions for $N$-body simulations with a multi-mass technique. We follow an approach based on \cite{Jun:2011jun} which is similar to that prescribed in \cite{Colin:2002nk,Robertson:2016qef,Drakos:2017gfy,2017MNRAS.470.4941H,Fischer:2020uxh}. Here the distribution ideally goes to zero at infinity. However, considering the finite extent of a halo, an exponential cut-off is introduced, $r_{\rm cut}$. The $r_{\rm cut}$ is identified as the maximum extension of the DM halo, which we have considered to be the virial radius. Following the prescription in reference \cite{Jun:2011jun,Burger:2018sqp}, we set the simulation parameters including the softening lengths $\epsilon$, force resolutions and the characteristic lengths $r_{\rm so}$ of the initial matter distribution. These parameters determine the characteristics of the individual simulated haloes. We list the values of these parameters in Table \ref{tab:Simulation}, for the NFW initial distribution. Relevant tests for this implementation are provided in \cite{Jun:2011jun} and we ensure that the code is able to reproduce them by optimizing the internal parameters like the scattering probability factor and  accuracy of time integration to ascertain the qualitative and quantitative validity of our simulations. All the cosmological parameters are set from the estimations of the PLANCK Collaboration data \cite{Ade:2015xua}. We have simulated haloes with six different masses in the range from $ 10^{10}M_{\odot} $ to $ 10^{15}  M_{\odot} $ for both the CDM scenario and the SIDM scenario with various scattering cross-sections using $10^6$ particles. The final density distribution is obtained after evolving the initial distribution of particles for $10$ Gyrs \cite{Miralda-Escude:2000tvu,Feng:2009hw,Kaplinghat:2015aga}. Admittedly, the resolution of our $N$-body simulation worsens as we move towards massive haloes. We therefore perform extensive convergence tests to demonstrate the stability of our simulations as sketched in appendix \ref{sec:appC}. We find that the fractional change in the matter density at the central region is less than $20\%$ for an order of magnitude change in the particle numbers. The results presented here should be interpreted modulo this resolution limit of the simulations. Each halo with the aforementioned masses have been simulated with self-interactions for eight different scattering cross-section per unit mass, set at $\sigma/m = (0.1\,, 0.2\,, 0.5\,, 2\,, 4\,, 6\,, 8\,, 10)\,{\rm cm}^2/{\rm gm} $. As we are simulating haloes with scattering cross-section per unit mass as large as $10\,{\rm cm}^2/{\rm gm} $, we have kept the evolution timestep to be considerably small by taking the scattering probability to be less than $0.1$, such that each particle can have at most one scattering per timestep \cite{Burkert:2000di,Jun:2011jun}. For comparison, we have additionally evolved all the haloes without self-interaction, which corresponds to the standard CDM scenario with larger values of $\sigma/m$ disfavored from \cite{Dave:2000ar,Rocha:2012jg,Kaplinghat:2013yxa}. For both the CDM and the SIDM scenarios, we use the density profiles that have achieved equilibrium during the period of $10$ Gyrs, i.e. the total energy of the system remains appreciably unchanged with the evolution time. We have simulated an ensemble of $4$ statistically independent realizations for every different combination of the halo mass and $\sigma/m$ considered in this work, in order to estimate the standard error in $N$ body simulations. We use these ensembles to compute the statistical fluctuations in the DM density distribution.


\begin{table}[t]
\centering   
\bigskip
\begin{tabular}{||p{4cm}|p{2.5cm}|p{3.5cm}|p{3cm}|}
    \hline
     $M_{\rm halo}$ ($10^{10}\,M_{\odot}$) & $r_{\rm cut}$ (kpc) & $r_{\rm so}$ (kpc) & $\epsilon$ (kpc) \\
     
\hline \hline
\setlength\arraycolsep{10pt}

$1$   & 43.7 & 0.43 & 0.02 \\ 

$10$  & 93.4 & 0.94 & 0.03 \\ 

$10^2$& 199  & 1.99 & 0.06 \\ 

$10^3$& 427  & 4.27 & 0.13 \\ 

$10^4$& 913  & 9.13 & 0.28 \\

$10^5$& 1952 & 19.52& 0.58 \\

\hline 
   
\end{tabular}
\caption{Parameters used for simulating the initial NFW particle distribution of the isolated haloes. \cite{Jun:2011jun}.}
\label{tab:Simulation}
\end{table}

\begin{figure*}[t]
\begin{center}
\subfloat[\label{sf:sm1}]{\includegraphics[scale=0.28]{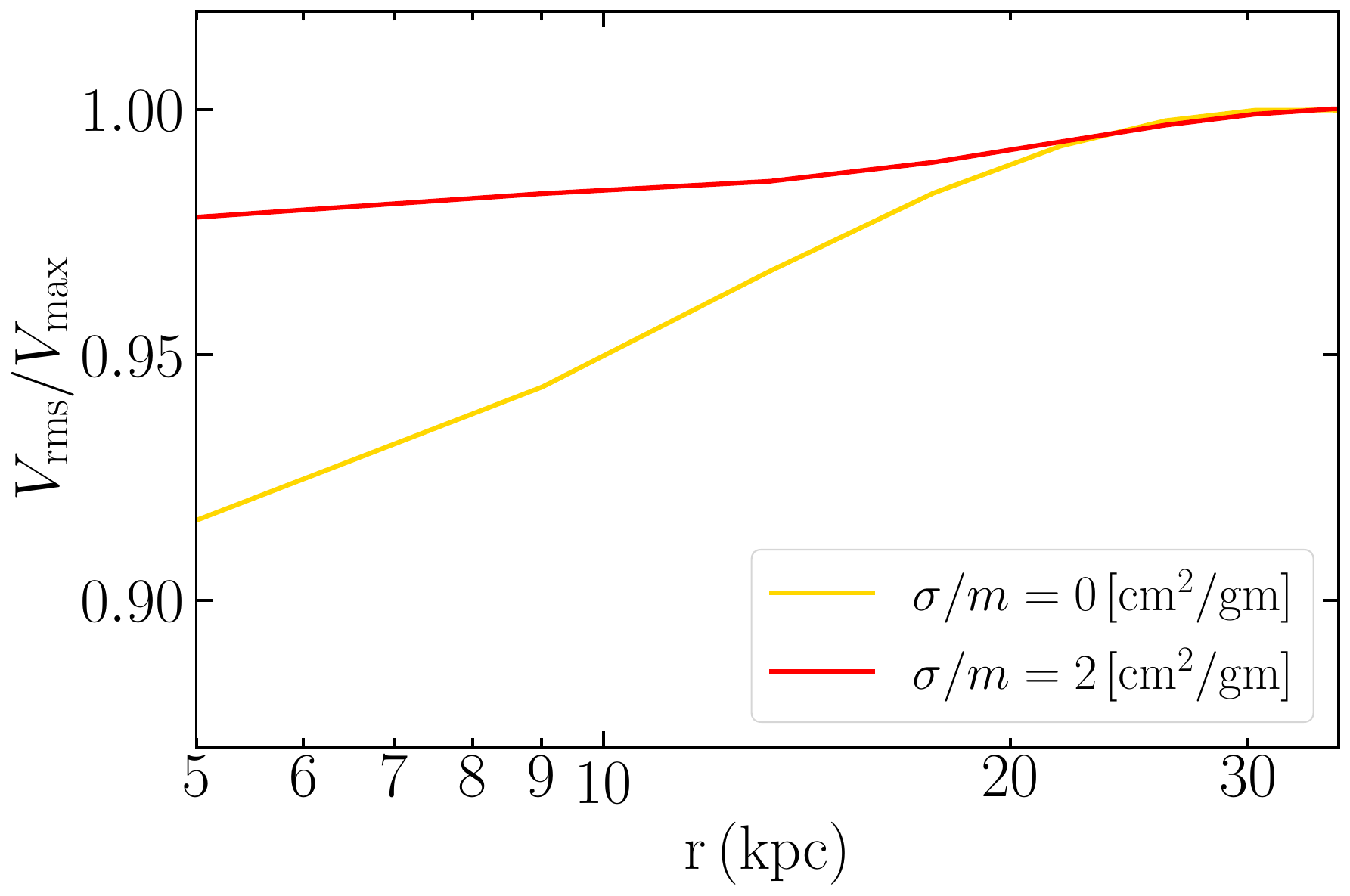}}
\subfloat[\label{sf:sm2}]{\includegraphics[scale=0.28]{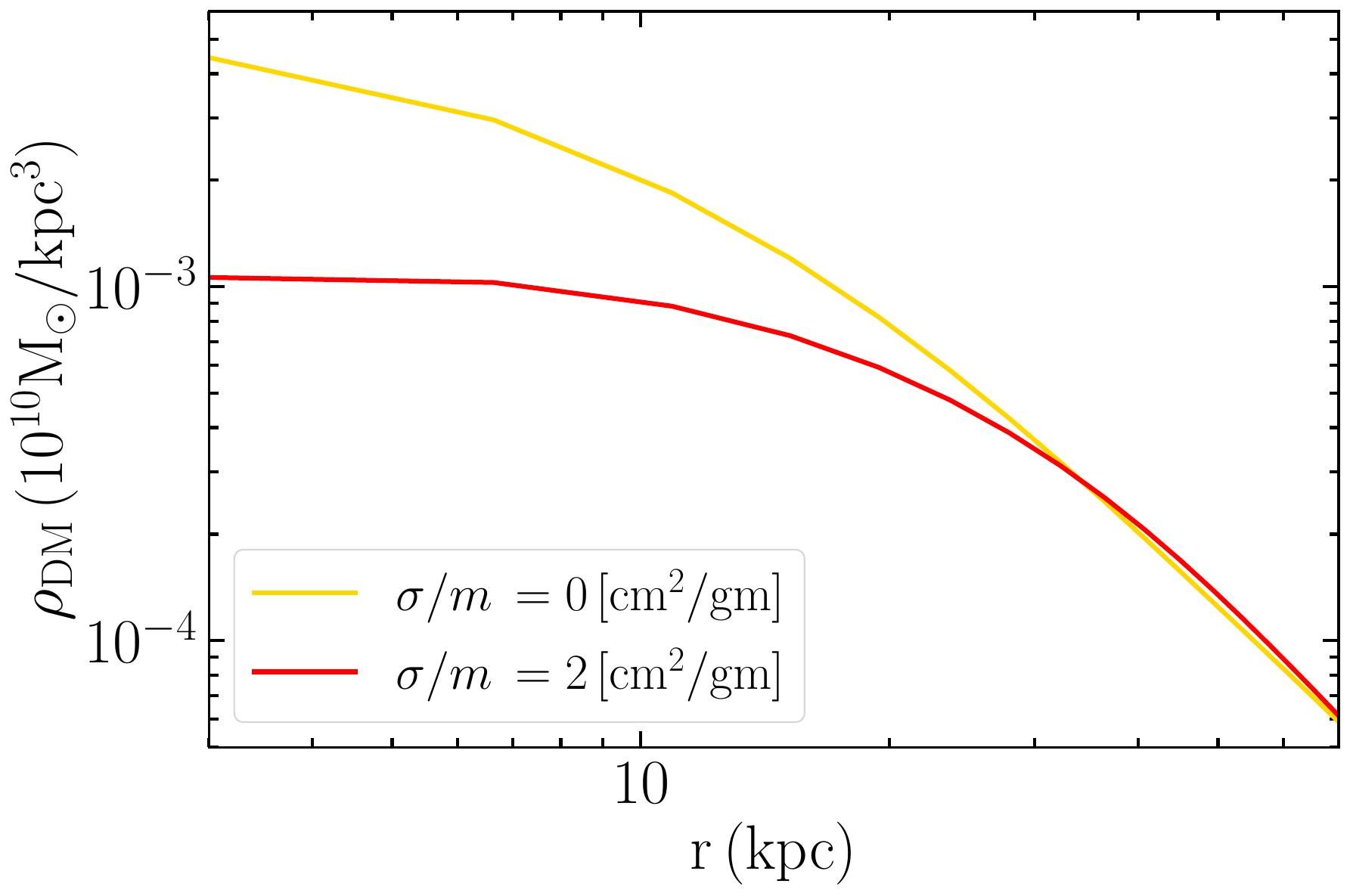}}
\caption{The left panel shows the variation of the velocity dispersion as a function of the radial distance from the halo center for the CDM (yellow) and the SIDM simulation (red) with $\sigma/m =2\, \rm cm^2 /\rm gm$ for the halo of mass $10^{13} M_{\odot}$. For the same halo, in the right panel we plot the density distribution for different values of scattering cross-section per unit mass.}
\label{fig:simulated}
\end{center}
\end{figure*}

The implications of DM self-interaction on the thermalisation of the core becomes apparent from the velocity  dispersion profile given in  figure \ref{sf:sm1}. We find that the velocity dispersion increases rapidly from the center towards the periphery for the CDM haloes, whereas the corresponding increment is comparatively softer for the SIDM haloes and remain almost constant at the center. This depicts the  expected scenario where the interior of the CDM haloes are gravitationally cooled, whereas DM self-interactions in the SIDM haloes thermalises the central region through the transfer of heat from the periphery leading to the formation of density cores inside the haloes. Our numerical estimates qualitatively agree with the results obtained in \cite{Burger:2018sqp,Banerjee:2019bjp} which considers high resolution cosmological simulations. As a sanity check to our implementation of self-interactions between DM, we find that the variation in density with radius shown in figure \ref{sf:sm2}, gradually decreases at the central region as the strength of self-scattering increases. This is also in accordance to the studies existing in literature \cite{Burger:2018sqp,Banerjee:2019bjp}.

\subsection{Core radius of the simulated haloes}
\label{sec:CoreSim}  
In order to estimate the core radius we fit the simulated haloes with density profiles  that incorporate a non-trivial core radius. Our analysis is based on the following two density profiles,

\begin{enumerate}
 
 \item \textbf{Semi-analytic distribution:} The semi-analytic distribution profile, also known as the Jeans model which was recently introduced in \cite{Kaplinghat:2013xca,Kaplinghat:2015aga}, has been argued to provide a better resemblance to $N$-body simulations with self-interaction \cite{Robertson:2020pxj}. The model considers a pseudo-isothermal profile within a characteristic radius ($r_1$) where self-interaction thermalises the core  and   beyond $r_1$ it follows the typical NFW distribution \cite{Navarro:1995iw} indicative of the collision-less nature at the low density periphery,
 \begin{equation}
\label{eq:jean}
\rho(r) = \left\{ \begin{array}{cc} \rho_{\rm iso}(r) \,\,\, = \, \rho_{\rm o} e^{-h(r/r_{\rm o})} \,  & \;\; r \leq r_{\rm 1} \\ \rho_{\rm NFW}(r)  = \frac{\rho_{\rm s}}{\frac{r}{r_{\rm s}} \left[1+\frac{r}{r_{\rm s}}\right]^2}\,  & \;\; \, r > r_{\rm 1}\, . \end{array} \right.
\end{equation}
The Jeans profile can be defined in terms of three independent parameters namely $r_{\rm o}$, $r_{\rm s}$ and $r_{\rm 1}$, using appropriate boundary conditions\cite{Valli:2017ktb}. Here the functional dependence of $h(r/r_{\rm o})$ on $r/r_{\rm o}$ has been adopted from \cite{Valli:2017ktb}. From definition the transition radius gives the estimate for the core size of a Jeans model involving self-interaction of DM.
 
\item \textbf{Cored NFW distribution:} A convenient parametrisation of a cored halo profile can be done in terms of a slightly modified version of the widely used NFW \cite{Lokas:2000mu,Lu:2005tu,10.1093/mnras/stz1698} density distribution, by introducing an additional free parameter $r_{\rm c}$  \cite{Newman_2013} which estimates the core radius. The distribution is given as
\begin{equation}
\label{eq:cNFW}
\rho_{\rm cNFW}(r)  = \frac{r_{\rm s} \rho_{\rm s}}{r_{\rm c} [1+\frac{r}{r_{\rm s}}]^2[1+\frac{r}{r_{\rm c}}]}\, ,
\end{equation}
where the symbols have their usual meanings. The profile implies a constant density core at $r < r_{\rm c}$, while retaining the usual NFW logarithmic slope of $-3$ at $r \gg r_{\rm c}$ . A similar parametrisation of the core radius as modification of the NFW distribution can be found in \cite{Read:2015sta}
 \end{enumerate}

\begin{figure*}[t]
\begin{center}
\includegraphics[scale=0.35]{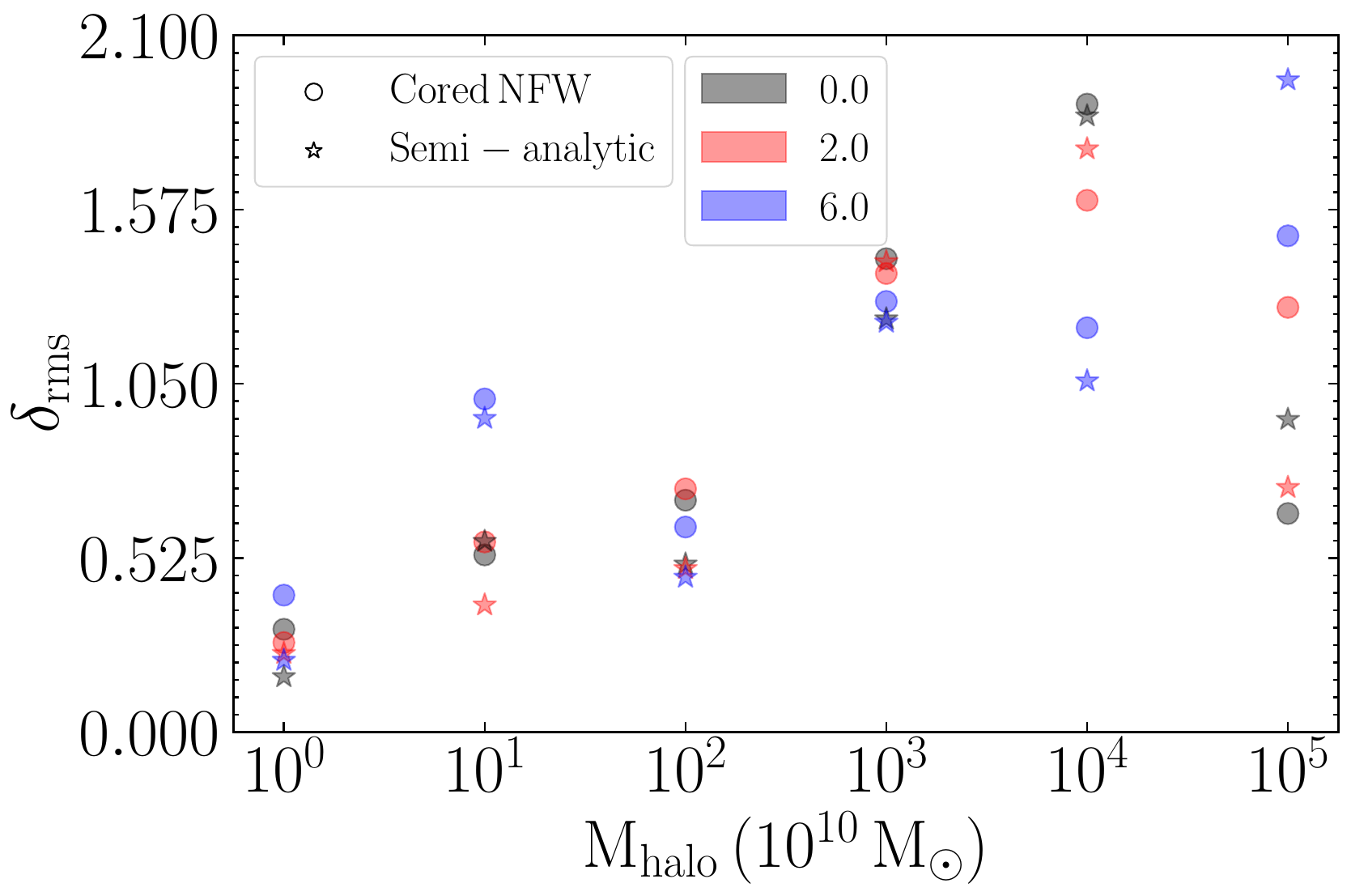}
\caption{Variations in the goodness-of-fit parameter, $\delta_{\rm rms}$ with the simulated DM halo mass for the cored NFW and semi-analytic distributions denoted by the circular and star shaped points respectively. This is shown for three sets of scattering cross-sections, namely the CDM (i.e. $\sigma/m=0 $), $2\, \rm cm^2/gm$ and $6 \, \rm cm^2/gm $ given by the grey, pink and blue points respectively.}
\label{fit-gdn}
\end{center}
\end{figure*}

We fit our simulated halo profiles with these models to extract the model dependent core radius. The goodness-of-fit for these density profiles with the simulated isolated haloes can be quantified in terms of $\delta_{\rm rms}$ given by
\begin{equation}
\label{eq:drms}
\delta_{\rm rms}^2 \equiv \frac{1}{N_{\rm bin}}\sum_{\rm i=1}^{\rm N_{\rm bin}}\left[\rm log_{10}\frac{\rho_{\rm data} ( r_i)}{\rho_{\rm mod}( r_i)}\right]^2, 
\end{equation}
which denotes the rms fluctuations of the obtained fits, with respect to the simulated data \cite{Robertson:2020pxj}. Here $N_{\rm bin}$ denotes the total number of radial bins in which the data is distributed, $\rho_{\rm data}$ is the average matter distribution corresponding to a particular bin obtained from the $N$-body simulation and $ \rho_{\rm mod} $ is the corresponding density fit for a particular distribution model. We have additionally considered the Burkert distribution\cite{Burkert:1995yz,Li:2020iib}, the Einasto distribution \cite{1965TrAlm...5...87E,Li:2020iib} and the core-modified Isothermal distribution \cite{10.1093/mnras/249.3.523,Li:2020iib}. However we limit our analysis only to the semi-analytic distribution and the cored NFW distributions, as they provide overall better fits to the observed and simulated DM density distributions, quantified through the goodness-of-fit estimates given by equation \eqref{eq:drms}. As can be seen from figure \ref{fit-gdn}, the fits are better for smaller galaxies that are observationally known to have larger relative core size.  Representative fits are shown in appendix \ref{sec:appA}. The core radius obtained from our simulated set of isolated haloes have been shown by the orange, red and brown curves in figure \ref{fig:MvsR}, for three sets of scattering cross-section i.e. CDM (i.e. $\sigma/m=0 $), $2$ and $6 \, \rm cm^2/gm $ respectively. 

\section{Distribution of core radius in the observed galaxies}
\label{sec:OBScore}

To compare the distribution of core radius from the simulated SIDM haloes with those derived from observations, we identify twelve haloes ranging from dwarfs to clusters with virial masses in the order of $ 10^{9} M_{\odot} $ to $ 10^{15} M_{\odot} $. These include three dwarf spheroidal galaxies [DD0-154, NGC-2366, IC-2754], six low surface brightness galaxies (LSB) [F-568-3, F563-V2, F563-1, NGC-3726, NGC3992, Malin-1] and three clusters [MS2137, A611, A2537]. The  details of the identified galaxies included in this study are given in Table \ref{tab:Galaxy}. From the observed rotation curve data of the chosen halos we identfy the DM contribution to the circular velocity using the following relation \cite{Read:2018fxs}.
\begin{equation}
\label{eq:vrms}
V_{\rm tot}^2(r)= v_{\rm DM}^2(r) + v_{\rm gas}^2(r) + v_{*}^2(r) .
\end{equation}
Here $V_{\rm tot}$ is the total galactic rotation curve, whereas $v_{\rm DM}$, $v_{\rm gas}$ and $v_{*}$ are the individual contribution of DM, gas and stellar contributions respectively. We consider a minimal disc setup for modeling the baryonic contribution of stars and galaxies from the corresponding references in Table \ref{tab:Galaxy}. The dwarf galaxies identified  are not satellites and they have mostly weak interactions with larger systems. This allows us to make systematic comparisons between the isolated halo simulations and observations. We extract the DM contribution to the rotation curves from the results of the THINGS survey \cite{deBlok:2008wp,Walter:2008wy}, that also report the contribution of the gas and baryons to the rotation curves supplemented by the Spitzer Infrared Nearby Galaxies Survey (SINGS) \cite{Kennicutt:2003dc} using equation \ref{eq:vrms}.

\begin{table}[t]
\centering   
\bigskip
\begin{tabular}{||p{3cm}|p{2.5cm}|p{3.5cm}|p{3.2cm}|p{1.5cm}||}
    \hline
    Galaxy Type & Galaxy Name & Halo mass ($10^{10}\,M_{\odot}$) & Properties & Reference \\
     
\hline \hline
\setlength\arraycolsep{10pt}

& NGC-2366 & 0.43 & Using HI rotation &  \cite{deBlok:2008wp, Oh_2011}\\ 

Dwarf Spheroidal & DD0-154  & 0.54 & curves (RC) & \cite{deBlok:2008wp, Oh_2011}\\ 

& IC-2754 & 1.46 & upto a few kpcs. &  \cite{deBlok:2008wp, Oh_2011}\\ 
\hline

& F-568-3 & 2.54 & Observed from H$\alpha$ &   \cite{KuziodeNaray:2006wh, KuziodeNaray:2007qi}\\ 

& F563-V2 & 1.65 & rotation curves & \cite{KuziodeNaray:2006wh, KuziodeNaray:2007qi}  \\

L.S.B & F563-1 & 2.8 &  upto $ \mathcal{O}(10)$ kpcs. & \cite{KuziodeNaray:2006wh, KuziodeNaray:2007qi}  \\

& NGC-3726 & 20.7  & Rotation curves of & \cite{article,Craciun:2020twu} \\  

& NGC-3992 & 37.2 &  HI distribution. & \cite{Bottema:2002yz,Craciun:2020twu} \\ 

& Malin-1 & 82 & Giant LSB, RCs of HI and H$\alpha$ observations & \cite{2010,2020} \\
\hline

& MS-2137 & 36307 & Observed lensing (Strong and &  \cite{Newman_2013, Newman_2013n}\\ 
Cluster & A-611 & 83176 &  Weak) measurements upto& \cite{Newman_2013n, Newman_2013}\\ 

& A-2537 & 218776 & a few Mpcs. & \cite{Newman_2013n, Newman_2013}\\ 
\hline 
   
\end{tabular}
\caption{Properties of the observed galaxies and clusters used in this study. Other relevant details can be obtained from \cite{Oman:2015xda}.}
\label{tab:Galaxy}
\end{table}

We infer the DM density profiles from the DM only rotation curves assuming a spherical DM distribution, using the following relation \cite{Oh_2011}
\begin{equation}
\label{eq:coll-t}
\rho(r)=\frac{1}{4 \pi G } \left[ 2\frac{V_{\rm c}}{r}\frac{\partial V_{\rm c}}{\partial r}+\left(\frac{V_{\rm c}}{r}\right)^2 \right].
\end{equation}
Here $V_{\rm c}(r)$ is the corresponding DM circular velocity at a radial distance $r$  from the halo center and $G$ is the universal Gravitational constant. For the  dwarf galaxies we use the masses given in \cite{Oh_2011}. The stellar  contribution in these galaxies is low, which reduces errors involving the uncertainty in the stellar mass-to-light ratio $\gamma_{*}$, giving a more accurate identification of the DM distribution. For the LSB galaxies we use the results of DensePak Integrated Velocity fields \cite{KuziodeNaray:2006wh, KuziodeNaray:2007qi, Bottema:2002yz} to extract the DM contribution to the rotation curves and the galaxy masses. For the galaxy clusters we use a sample of massive, relaxed galaxy clusters with centrally-located brightest cluster galaxies (BCG) at redshifts z $= 0.2 - 0.3$, derived from observational tools of strong and weak gravitational lensing combined with resolved stellar kinematics within the BCGs'. The total radial density profile with both DM and baryons over scales of $ 3-3000$ kpc  are presented in \cite{Newman_2013n, Newman_2013}. Our estimated core radius for the observed set of galaxies and clusters are represented by the magenta line in figure \ref{fig:MvsR}. The corresponding shaded bands denote the region of uncertainty associated with the core estimates from observations. The individual fits to the inferred DM density distribution are shown in appendix \ref{sec:appB}.

\section{Constraints on DM self-interaction from the distribution of core size in isolated haloes}
\label{sec:results}

\begin{figure*}
\begin{center}
\subfloat[\label{sf:MR-KP}]{\includegraphics[scale=0.3]{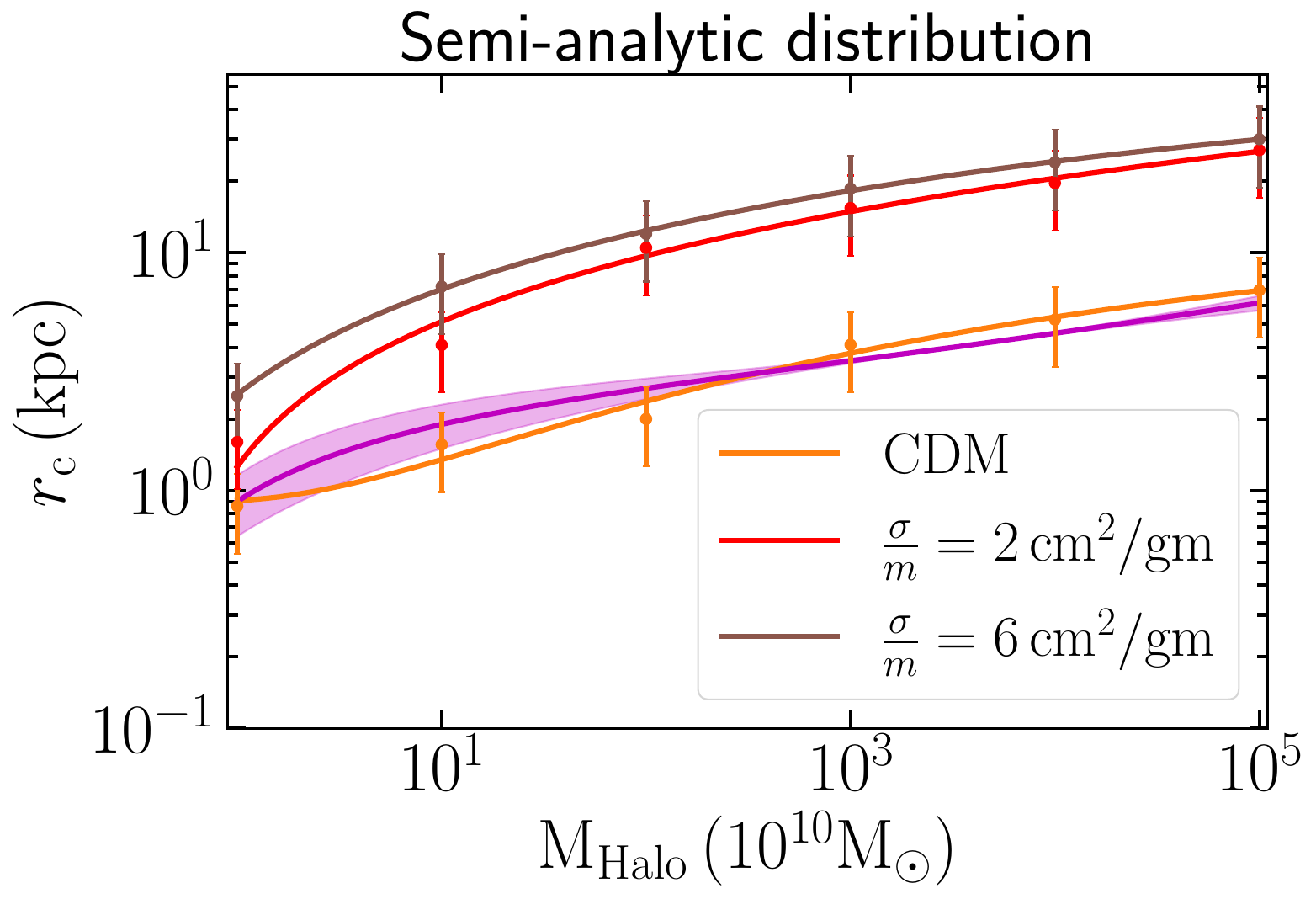}}
\subfloat[\label{sf:MR-CN}]{\includegraphics[scale=0.3]{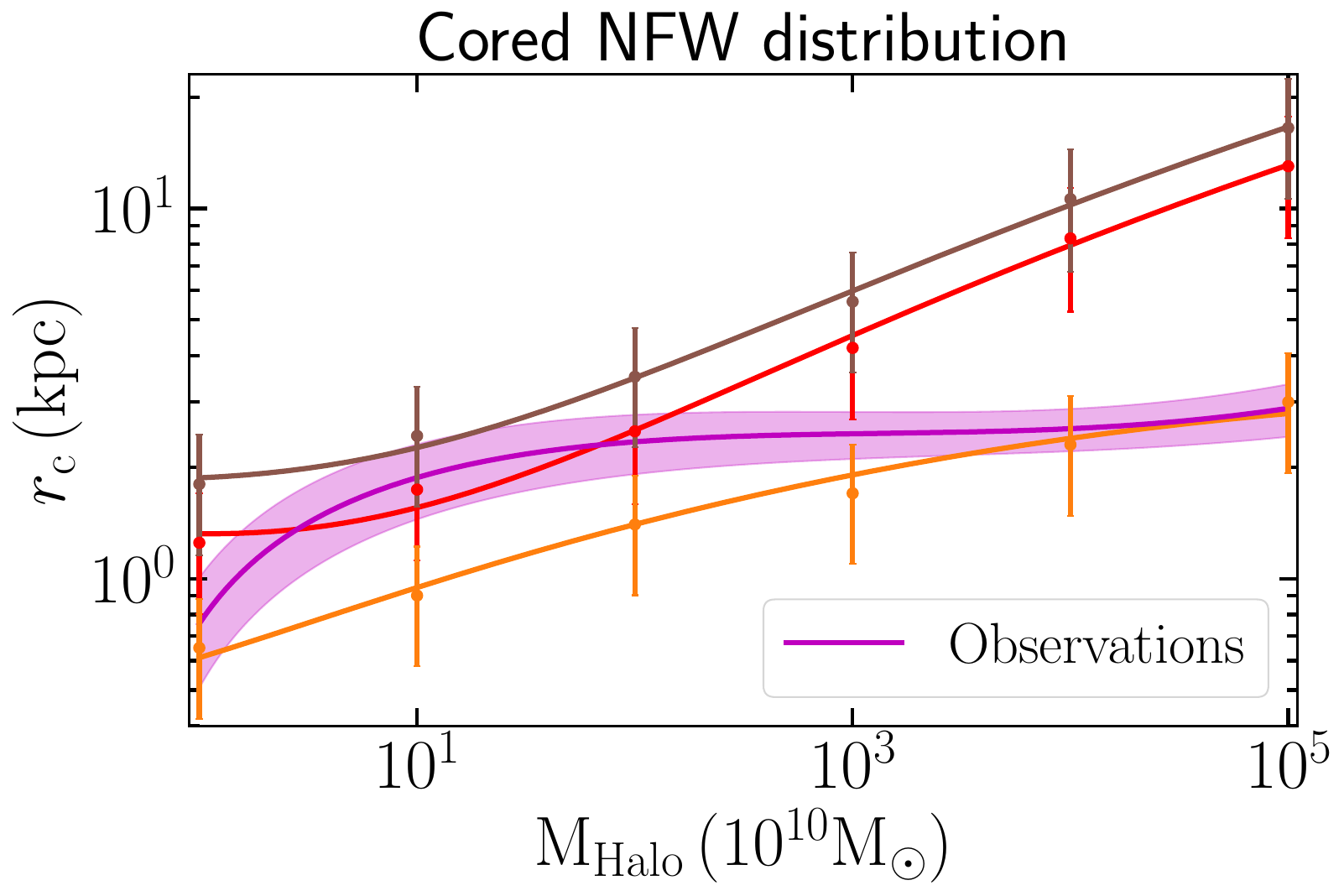}}
\caption{Variation of estimated core radius with the DM halo mass. The left and right panel denotes the variations for the semi-analytic distribution and the cored NFW distribution respectively. The orange, red and brown points represent the scattering cross-sections per unit mass from our simulations for CDM, $\sigma/m = 2\ \rm cm^2 /\rm gm$ and $\sigma/m =6\ \rm cm^2 /\rm gm$ respectively. The corresponding lines represent the best fit curves to the respective colored points. The magenta lines denotes the fit to the DM halo core inferred from  observed data.} 
\label{fig:MvsR}
\end{center}
\end{figure*}

We obtain constraints on the DM self-interaction by statistically comparing the core radius distributions extracted  from  observations with the ones obtained by our simulated SIDM haloes. To obtain the best fit core radius we utilize standard regression tools like the {\tt scipy} \cite {Virtanen:2019joe} module in python and inbuilt functions in Mathematica. These comparisons and hence the obtained constraints are sensitive to the choice of the DM density profile. The core radius distributions have been presented in figure \ref{fig:MvsR}, where the orange, red and brown points give the estimates of core radius from $N$-body simulations with zero, $ 2 $ and $6\, \rm cm^2 /\rm gm $ interaction strengths respectively as a function of the DM halo mass. The corresponding vertical lines give an estimate of the uncertainties involved in determining $r_{\rm c}$, that has been obtained by simulating four different realizations of every scenario and extracting the statistical errors in our simulations. The polynomial fits to these points are shown by the corresponding coloured curves. We are now in a position to compare our observed and simulated cores and measure the goodness-of-fit between the two.

\begin{figure*}[t]
\begin{center}
\subfloat[\label{sf:Chi-KP}]{\includegraphics[scale=0.253]{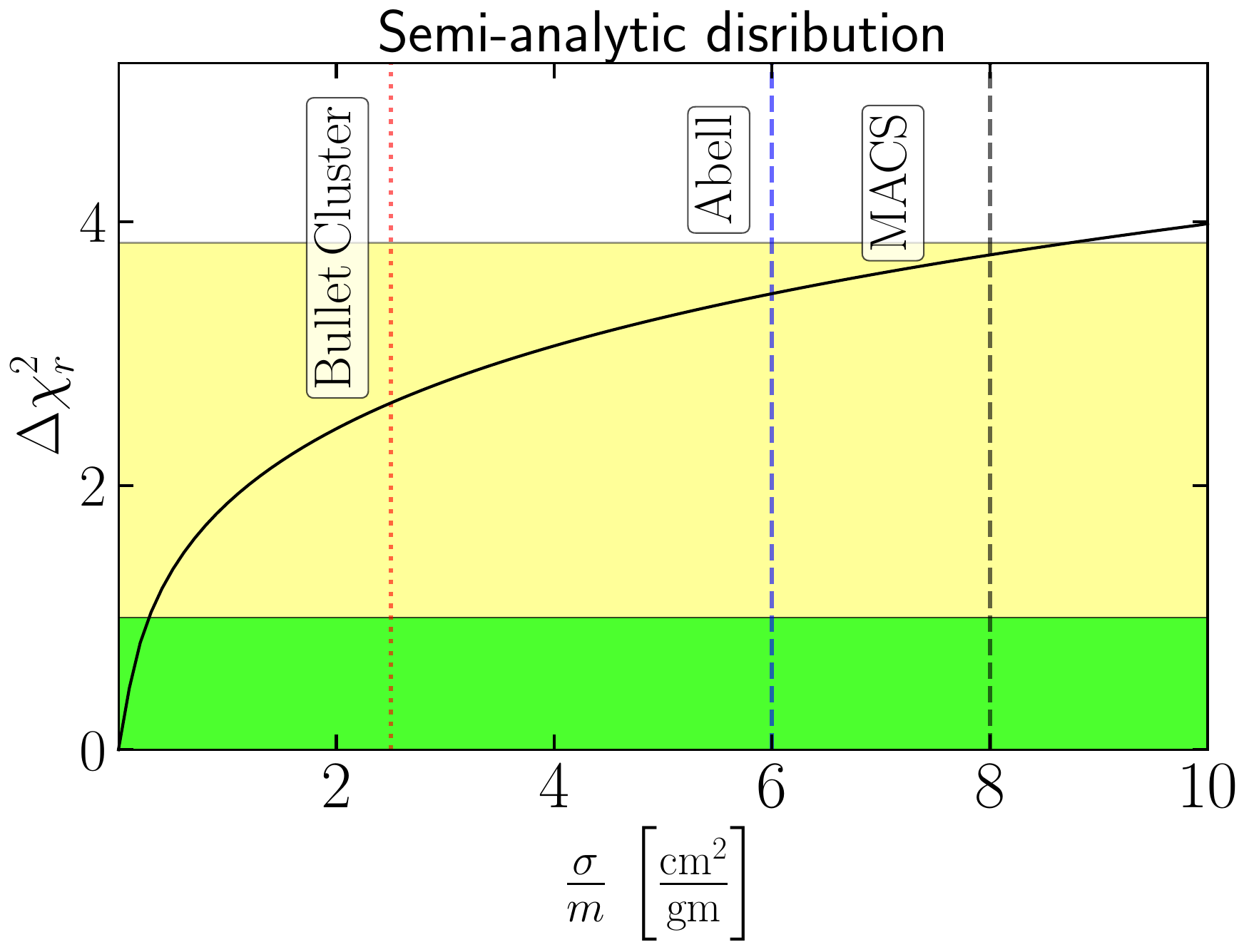}}
\subfloat[\label{sf:Chi-CN}]{\includegraphics[scale=0.27]{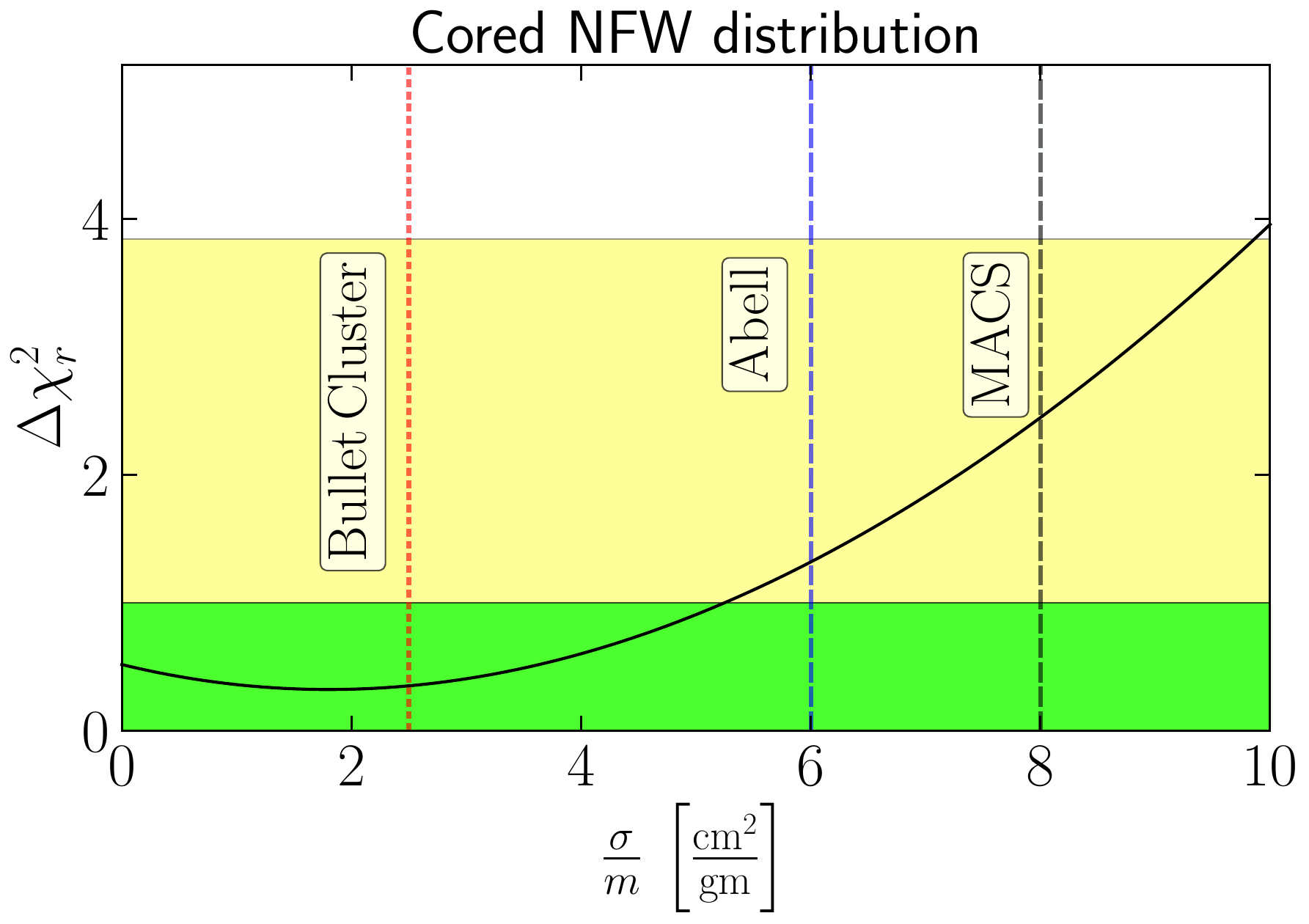}}
\caption{Effective $\chi_{\rm r}^2$ as a function of $\sigma/m $. The left and right panels denote the $\Delta \chi_{\rm r}^2$ variations for the semi-analytic distribution and the cored NFW distribution respectively. The green and yellow regions denote the $68 \%$ and $95 \%$ confidence regions respectively with respect to the minimum $\chi_{\rm r}^2$. The red, blue and brown dashed vertical lines represent the upper limits at $95\%$ confidence obtained from the Bullet cluster, Abell Cluster and MACSJ0025 (Baby bullet) cluster respectively.} 
\label{fig:chivar}
\end{center}
\end{figure*}

The distribution of the core radius provides us with a handle to statistically compare the DM self-interaction to obtain limits on the interaction cross-section. This approach of comparing the core radius distribution rather than comparing the core radius of individual isolated haloes, make the analysis more robust by reducing the simulation errors. The goodness-of-fit for the distribution is done using the standard chi-square $\chi_{\rm r}^2$,
\begin{equation}
\label{eq:chi}
\chi_{\rm r}^2=\frac{1}{N}\sum_{\rm i}^{\rm N}\left[\frac{(r_{\rm c}^{\rm s})_{\rm i}-(r_{\rm c}^{\rm o})_{\rm i}}{\sigma_{\rm i}}\right]^2, 
\end{equation}
and the reduced chi-square defined by $\Delta \chi_{\rm r}^2=\chi_{\rm r}^2 - \left(\chi^2_{\rm r}\right)_{\rm min},$ where $N$ corresponds to the total number of points  sampled from the best fit curves, $r_{\rm c}^{\rm s}$ and $r_{\rm c}^{\rm o}$ are the core radius obtained from the simulations and observational data respectively and $\sigma_{\rm i}$ are the errors associated with the simulations. For a given model of density  profile we use the maximum simulation error from the simulated haloes. As discussed in section \ref{sec:CoreSim}, we restrict our comparative analysis between observations and simulations to the semi-analytic distribution and cored the NFW distribution. The estimated $ \Delta \chi_{\rm r}^2$ as a function of the interaction cross-section per unit mass for the two models are shown in figure \ref{fig:chivar}. The plots in figure \ref{fig:chivar} indicate that the overall fit to the semi-analytic distribution minimizes at $\sigma/m =0$, whereas the cored NFW analysis prefers a small scattering cross-section of $1.7 \, \rm cm^{2}/ \rm gm $ where the $ \chi_{\rm r}^2$ minimizes.

The green and yellow shaded regions in figure \ref{fig:chivar} represent the $68\%$ and $95\%$ confidence band of our chi-square distribution \cite{ParticleDataGroup:2020ssz}. We report $\sigma/m $ below $8.6\,(9.8) \, \rm cm^{2}/ \rm gm $ to lie within $95 \%$ confidence level and $0.3\,(5.4) \, \rm cm^{2}/ \rm gm $ within $68 \%$ confidence level for the Jeans (cored NFW) model. The bounds on SIDM interaction strengths from the  Bullet cluster \cite{Harvey:2015hha, Randall:2007ph, Robertson:2016xjh}, the Ablle Cluster \cite{Kahlhoefer:2015vua} and the MACSJ0025 cluster (baby bullet) \cite{Brada__2008}   have been shown by the vertical dashed lines in figure \ref{fig:chivar}. The limits obtained from this analysis are  consistent with the aforementioned limits and other studies quoted in the literature \cite{Bernal:2019uqr, Ackermann:2013yva, Massey:2015dkw, Modak:2015npa}.

These  bounds are  premised on the assumption that the distribution of DM is spherically symmetric in the isolated halos and have an isotropic DM self-interaction. Any impact of non-sphericity or triaxiality  \cite{Oman:2017vkl, Jing:2002np} has been neglected in the results presented here. Inclusion of baryons provides a competing mechanism to thermalise the central region of the halos driving core formation. Additional effects like supernova feedback, stellar formation, viscosity etc would further aid in the formation of cores. Expectedly the inclusion of these phenomenon would make the bounds in the DM self-interaction more stringent. The limits originating from the DM only simulations  can be considered to be more conservative and the results presented here should be interpreted within this context. 

\section{Conclusion}
\label{sec:conc}

In this work we study the feasibility of using the distribution of core radius in isolated haloes as a mean of constraining DM self-interaction. As a proof of principle we simulate haloes  with six different masses in the mass range  $10^{10}- 10^{15}$ $M_{\odot}$ for the CDM as well as the SIDM scenario, having $\sigma/m $ in the range of $ (0-10) \, \rm cm^{2}/ \rm gm $. We thereby compare these simulations with observational data for twelve different haloes in the same mass range. The comparison requires us to assume an underlying model for the density profile making the translated  limits on the self-interaction cross-section sensitive to these choices. However a model independent definition of the halo core radius would make the study more robust. A possibility is to utilize the discontinuous change in the slope of the rotation curves between the core and the periphery in order to identify the core radius directly from the density profiles. A detailed study on the numerical feasibility and stability of this is beyond the mandate of our present work. For a velocity independent point like interactions we find that for the semi-analytic distribution, the CDM scenario  provides an overall best fit to the wide range of halo mass comprising of dwarfs, LSBs and clusters. Whereas the cored NFW model prefers a small but non-zero scattering cross-section. Remaining within the limits of spherical symmetry and employing DM only simulations, we obtain a conservative bound on $\sigma/m $ to be below  $ 9.8 \ \rm cm^{2}/ \rm gm $ at $95\%$ confidence level. This is comparable to bounds from galaxy cluster mergers available in the  literature. However these bounds should be interpreted carefully as they are sensitive to the uncertainties associated with the simulations and observations.

Distribution of core sizes in the isolated haloes may provide a complementary handle to compare and constrain DM self-interaction. Here we demonstrate its utility in the simplest context of velocity independent point like interactions. However, higher resolution cosmological simulations from which a large number of isolated haloes may be identified can be employed to compare and distinguish between the strength and velocity dispersion of DM self-interaction. Such dedicated studies are now in order and will be carried out elsewhere.

\paragraph*{Acknowledgements\,:} 

The authors would like to thank Somnath Bharadwaj, Nishikanta Khandai, Tarak Nath Maity, Avik Banerjee and Debajit Bose for the helpful discussions and useful comments. TSR acknowledges the hospitality of ICTP, Trieste under their associateship program towards the completion of this project. SS acknowledges the University Grants Commission (UGC) of the Government of India for providing financial assistance through Senior Research Fellowship (SRF) with reference ID: 522157. AKS acknowledges the support of the Israel Science Foundation (grant no. 255/18). The authors also acknowledge the National Supercomputing Mission (NSM) for providing computing resources of ‘PARAM Shakti’ at IIT Kharagpur, which is implemented by C-DAC and supported by the Ministry of Electronics and Information Technology (MeitY) and Department of Science and Technology (DST), Government of India.

\appendix
\FloatBarrier
\section{Fits to our simulated data}
\label{sec:appA}

\begin{figure}[H]
	\begin{center}
		\subfloat[\label{sf:Ms1}]{\includegraphics[scale=0.18]{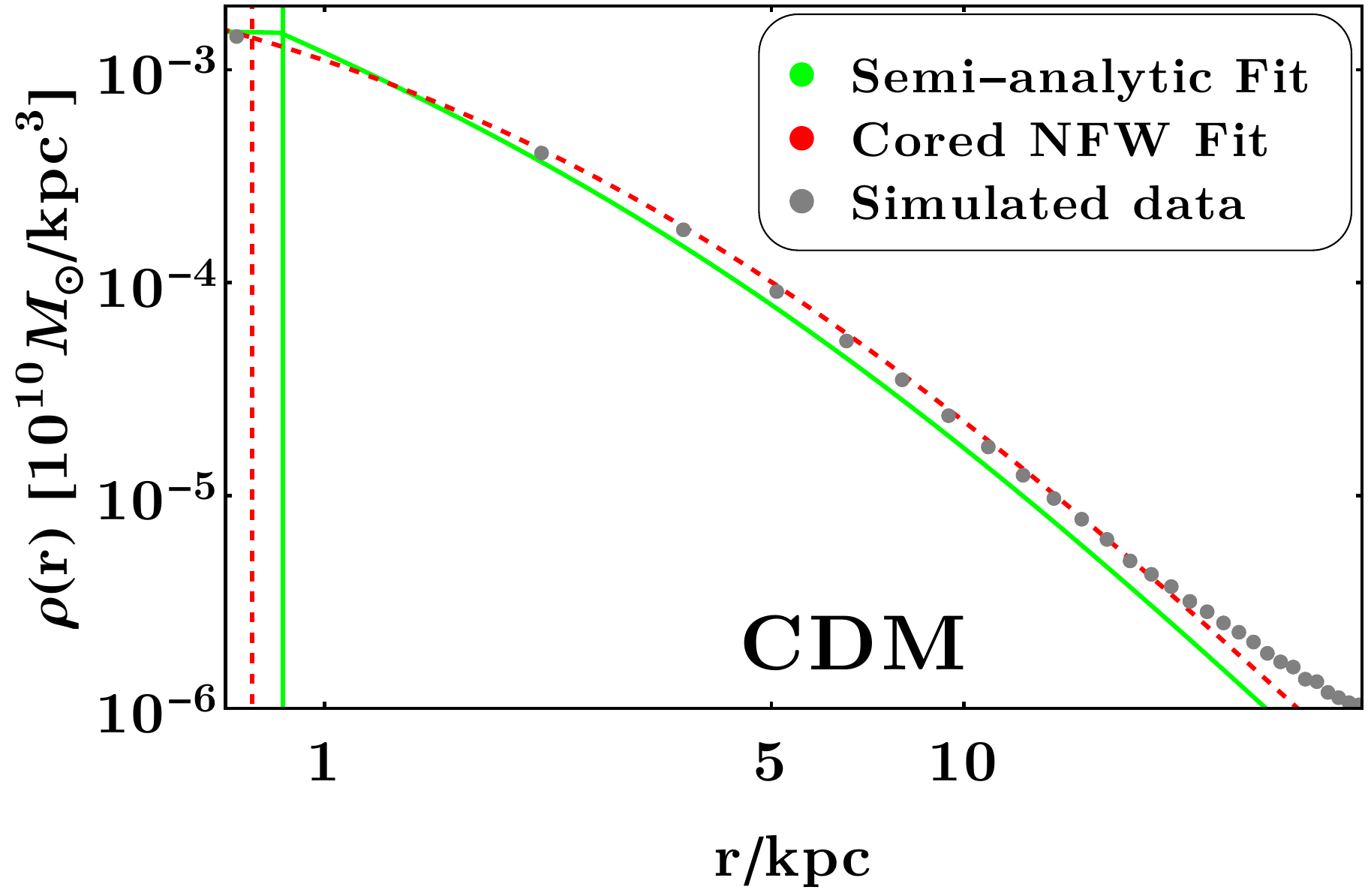}}
		\subfloat[\label{sf:Ms2}]{\includegraphics[scale=0.18]{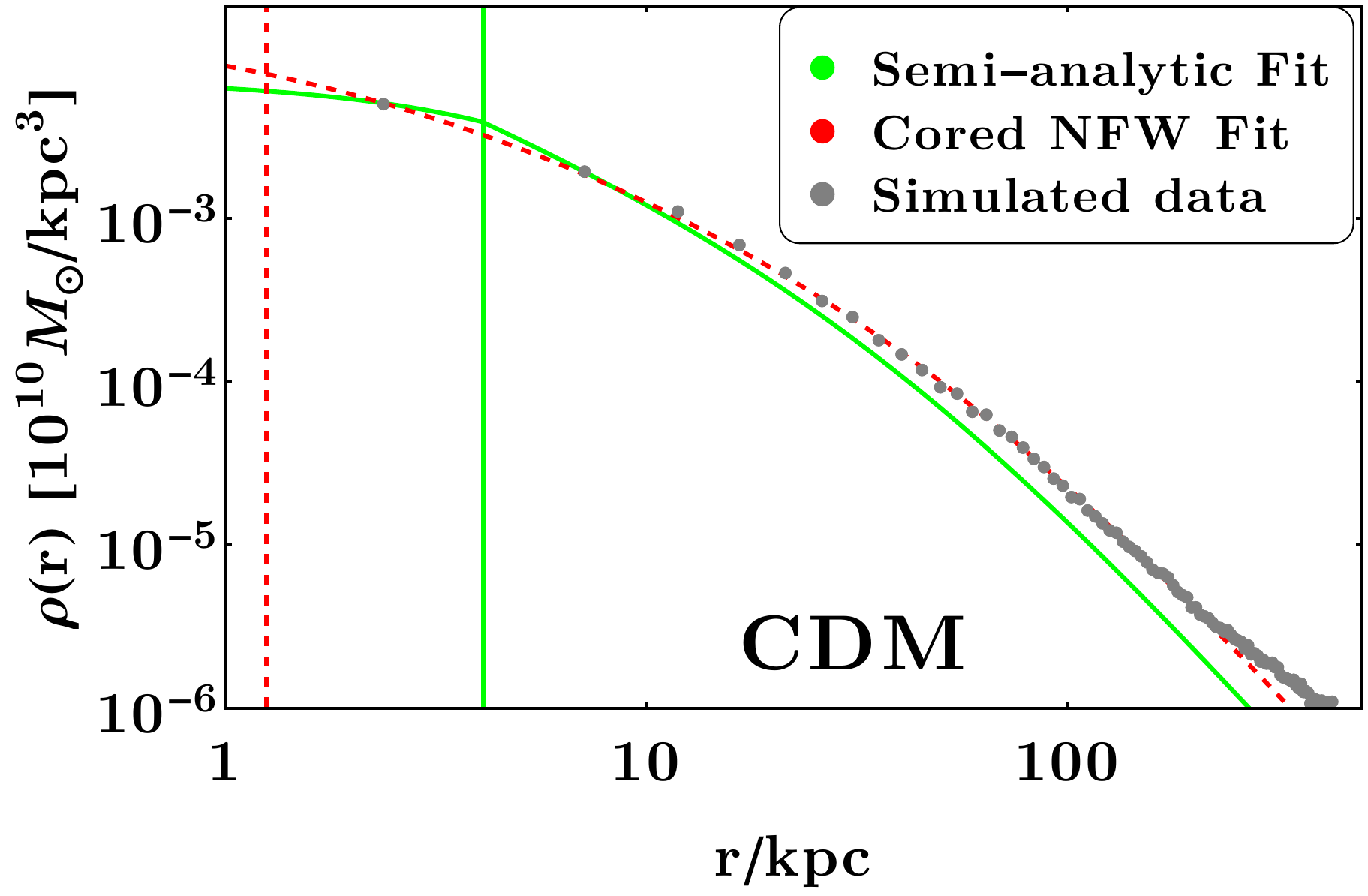}}
		\subfloat[\label{sf:Ms3}]{\includegraphics[scale=0.18]{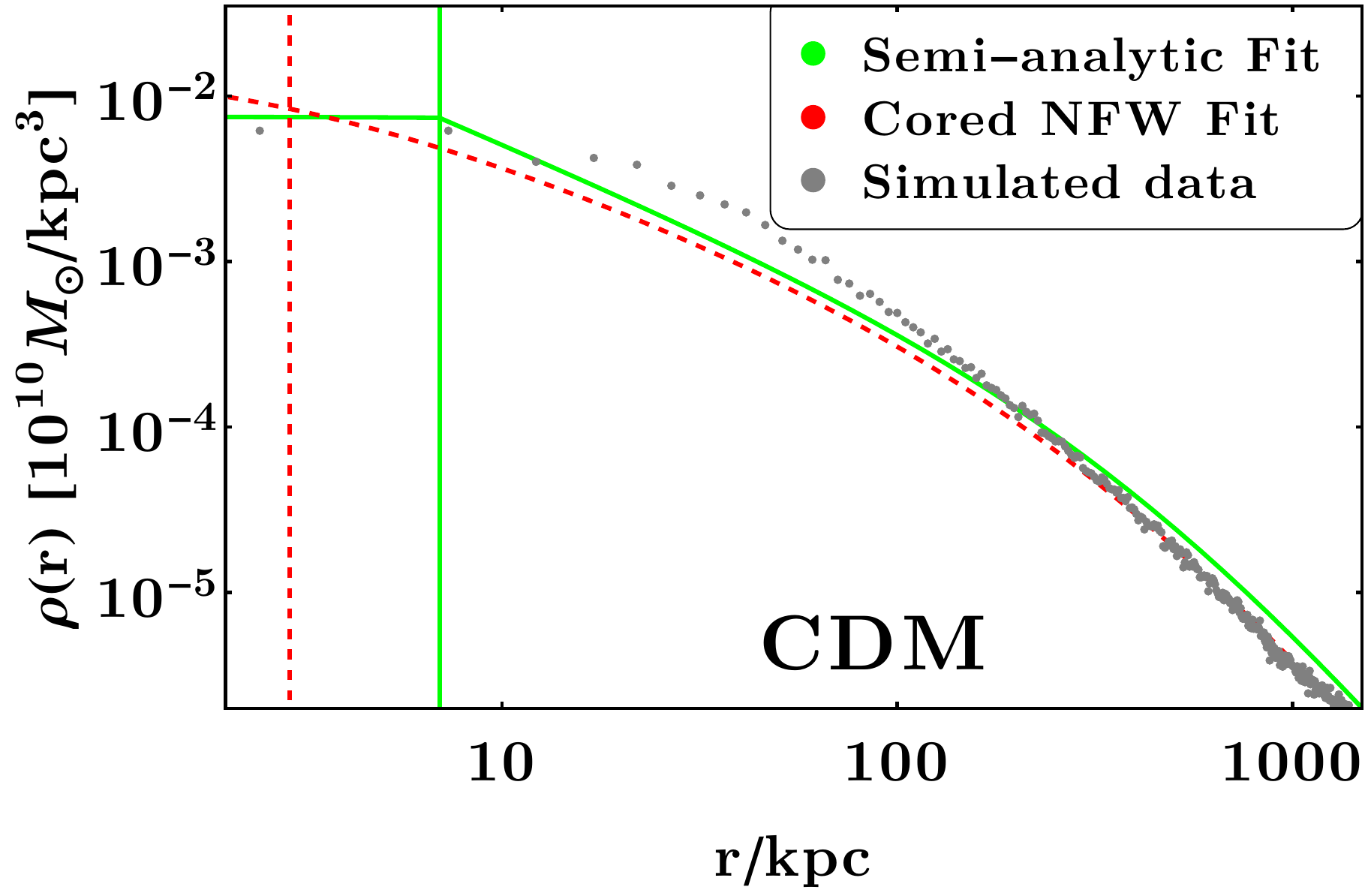}}
		\linebreak
		\subfloat[\label{sf:Ms4}]{\includegraphics[scale=0.18]{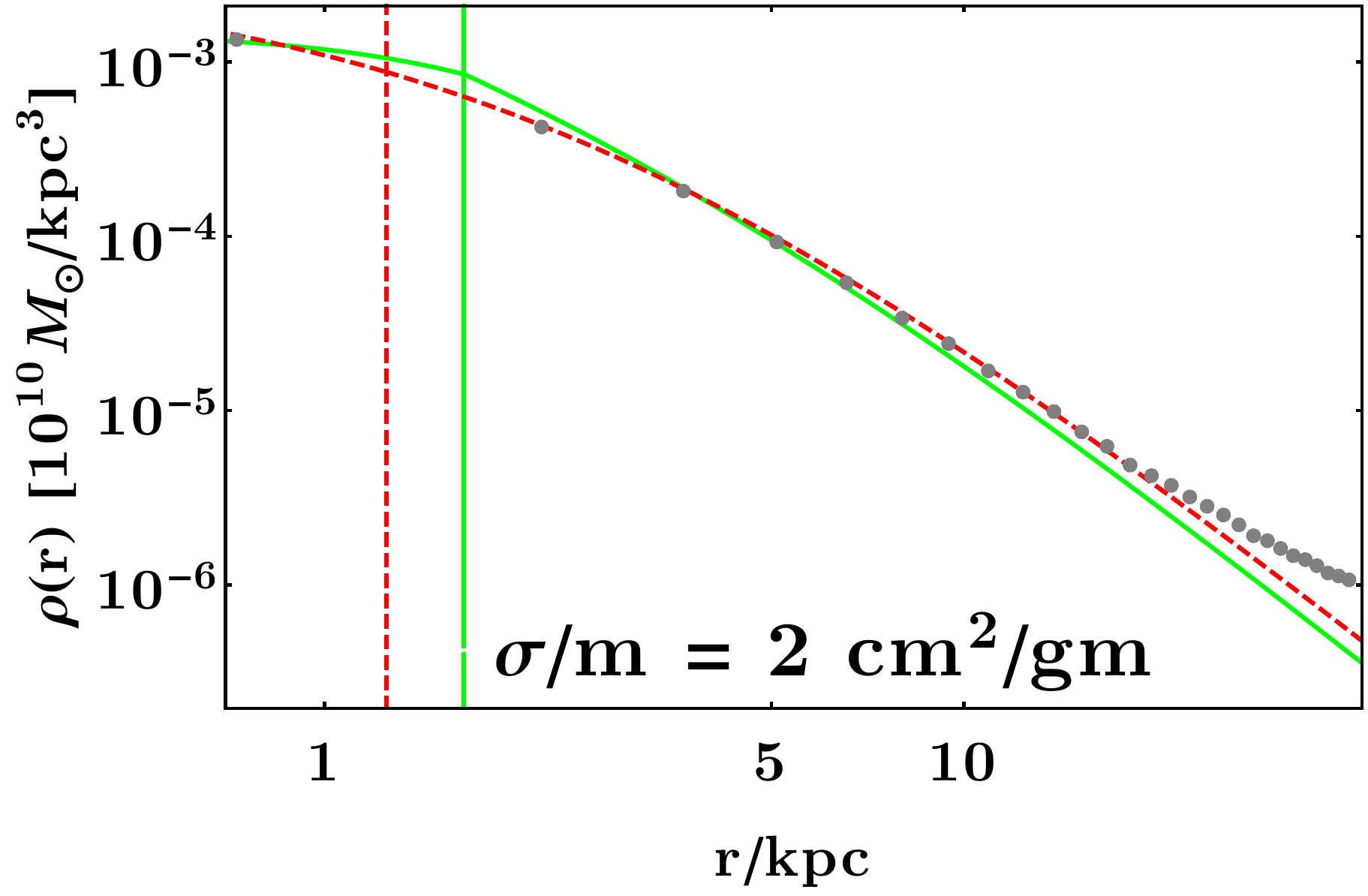}}
		\subfloat[\label{sf:Ms5}]{\includegraphics[scale=0.18]{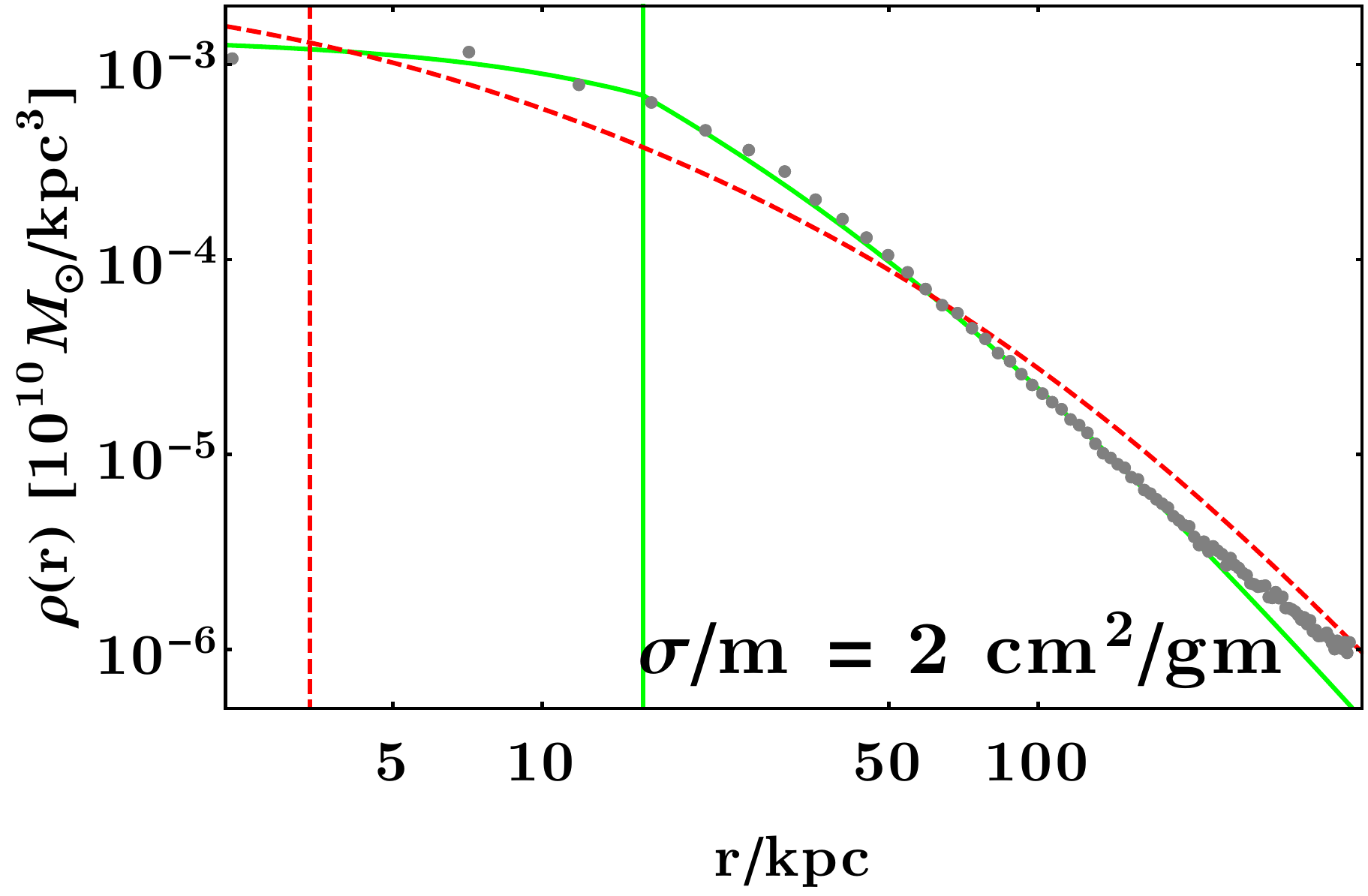}}
		\subfloat[\label{sf:Ms6}]{\includegraphics[scale=0.18]{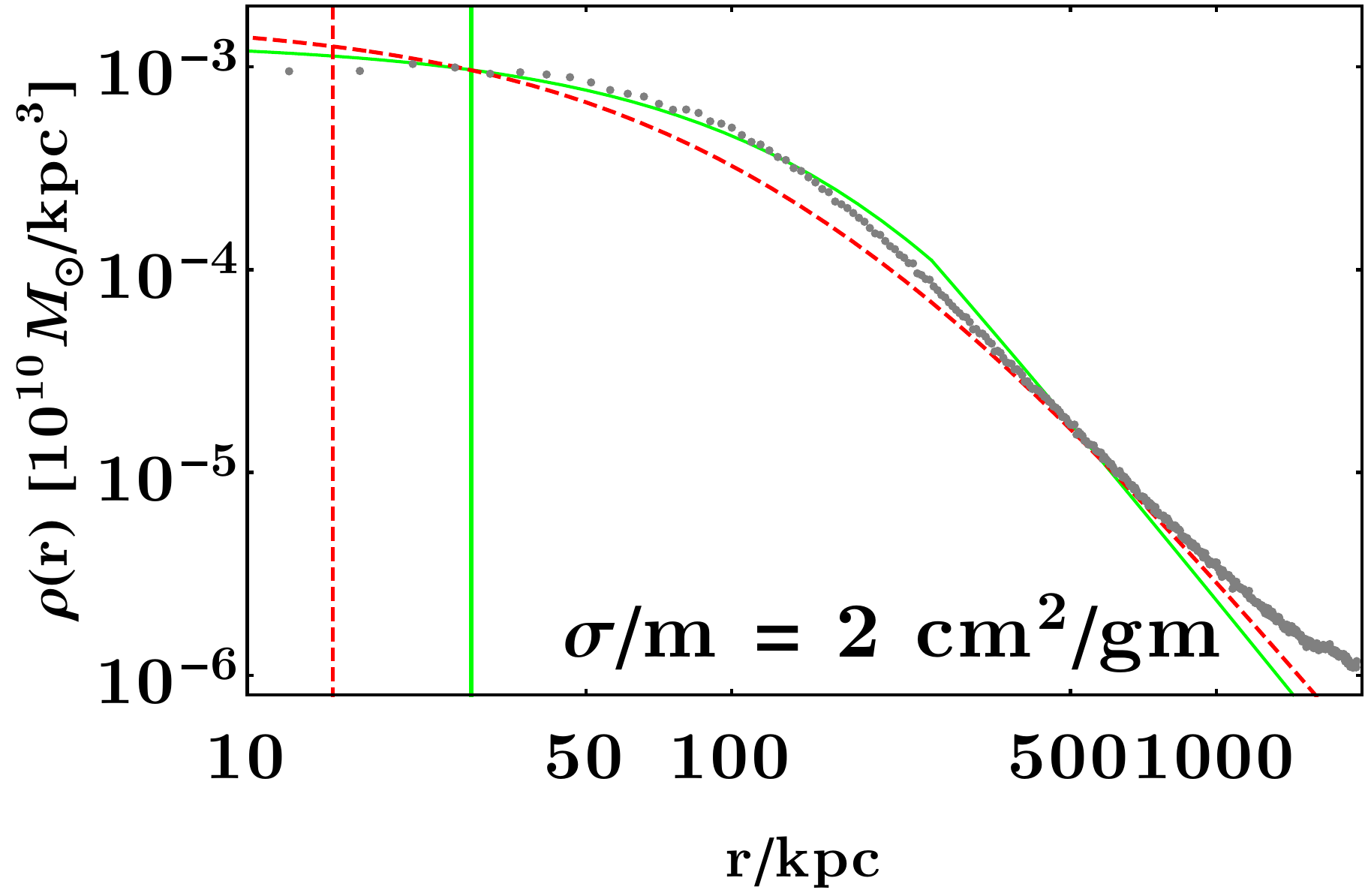}}
		\caption{Model fits to the DM distributions obtained from the $N$-body simulations.}
		\label{fig:appA}
	\end{center}
\end{figure}

As discussed in section \ref{sec:Nbodyint}, we show here the fits to the individual density distributions obtained for our $N$-body simulations. The density distributions for the CDM scenario and a SIDM scenario with $\sigma/m = 2 \, \rm cm^2/gm $ are shown in the top and bottom panel of figure \ref{fig:appA} respectively. This is shown for three different halo mass namely $10^{10}M_{\odot}$, $10^{13}M_{\odot}$ and $10^{15}M_{\odot}$ in the left, middle and right panels of the figure respectively. The green solid and red dashed curves represent the best fit curves for the semi-analytic distribution and the cored NFW distribution respectively, the corresponding vertical colored lines denote the extent of its core radius. The simulated data are shown by the gray points.
\FloatBarrier
\section{Fits to the observational data}
\label{sec:appB}

\begin{figure*}[h]
\begin{center}
\subfloat[\label{sf:Fd1}]{\includegraphics[scale=0.147]{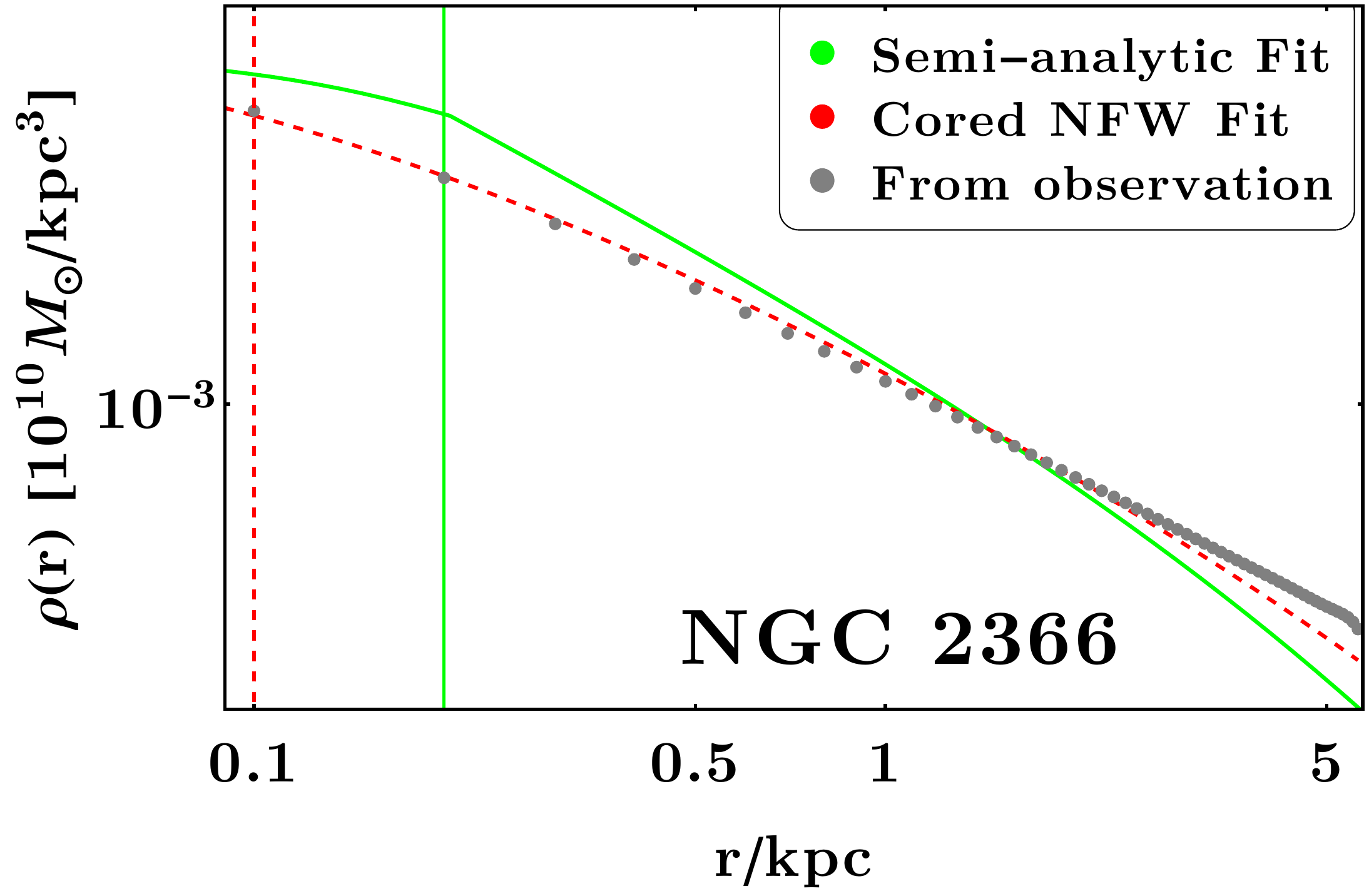}}
\subfloat[\label{sf:Fd2}]{\includegraphics[scale=0.147]{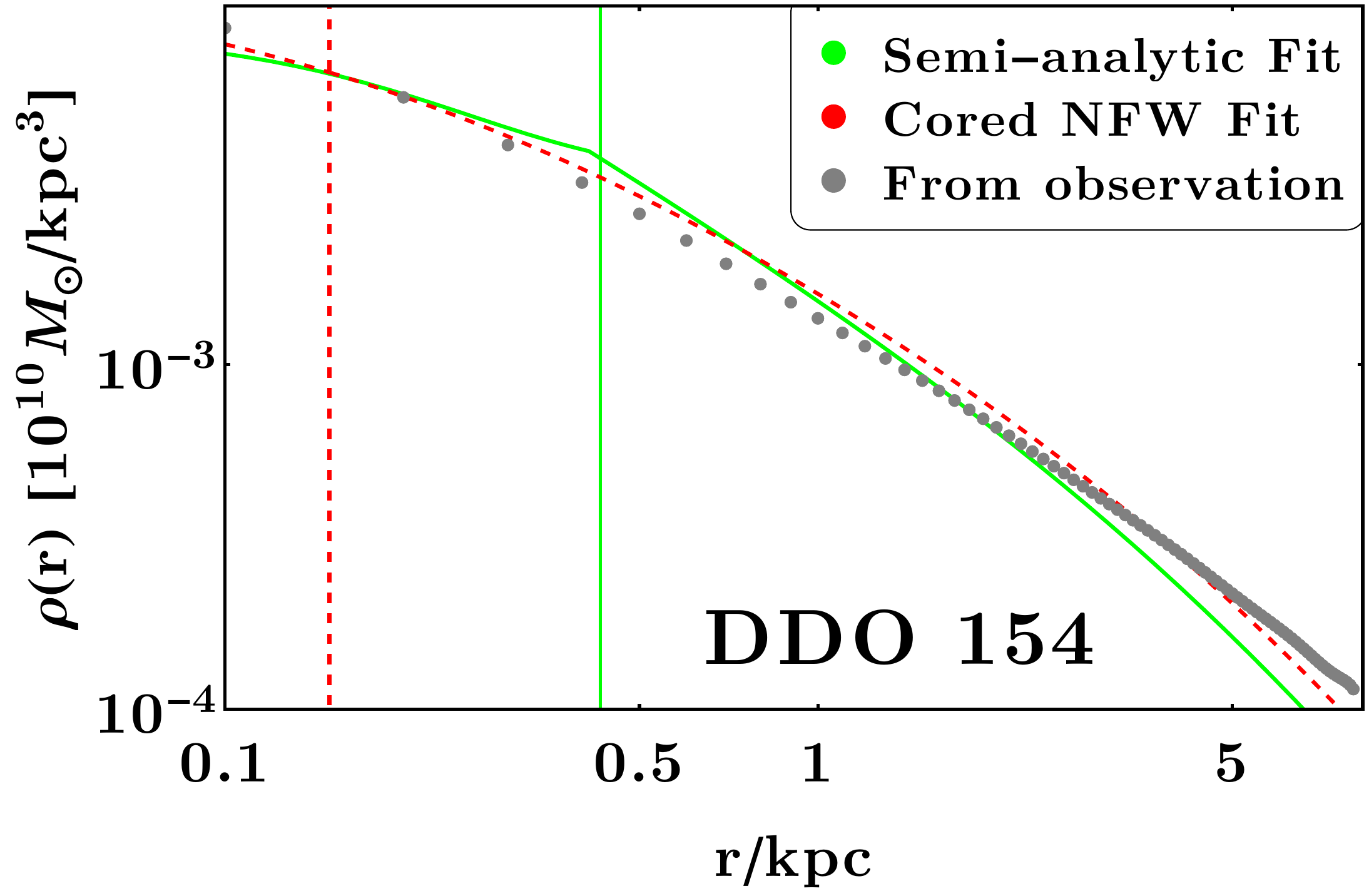}}
\subfloat[\label{sf:Fd3}]{\includegraphics[scale=0.147]{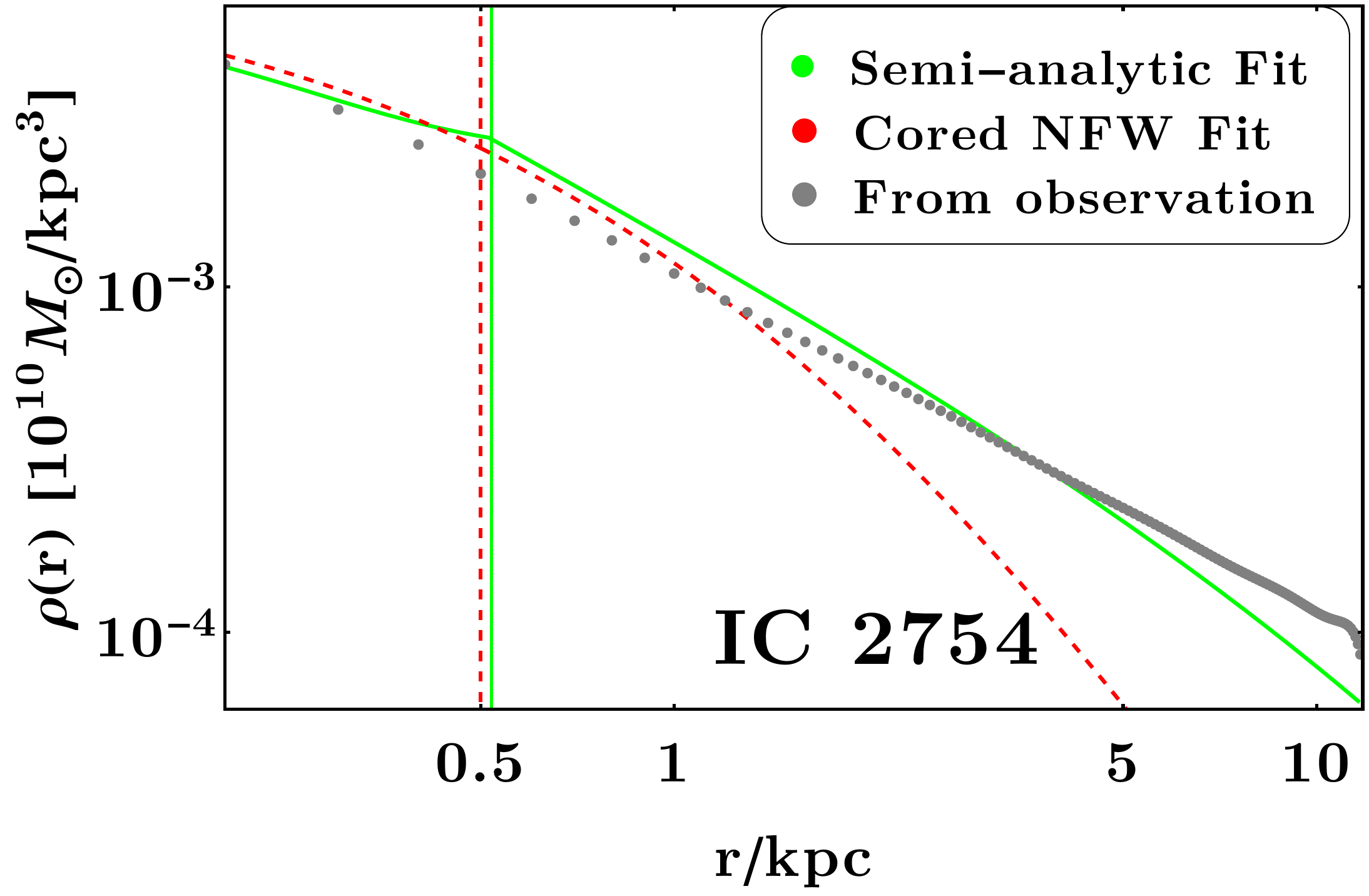}}
\caption{Model fits to the DM distributions inferred from dwarf galaxy observations.}
\label{fig:appB1}
\end{center}
\end{figure*}

\begin{figure*}[h]
\begin{center}
\subfloat[\label{sf:Fl1}]{\includegraphics[scale=0.147]{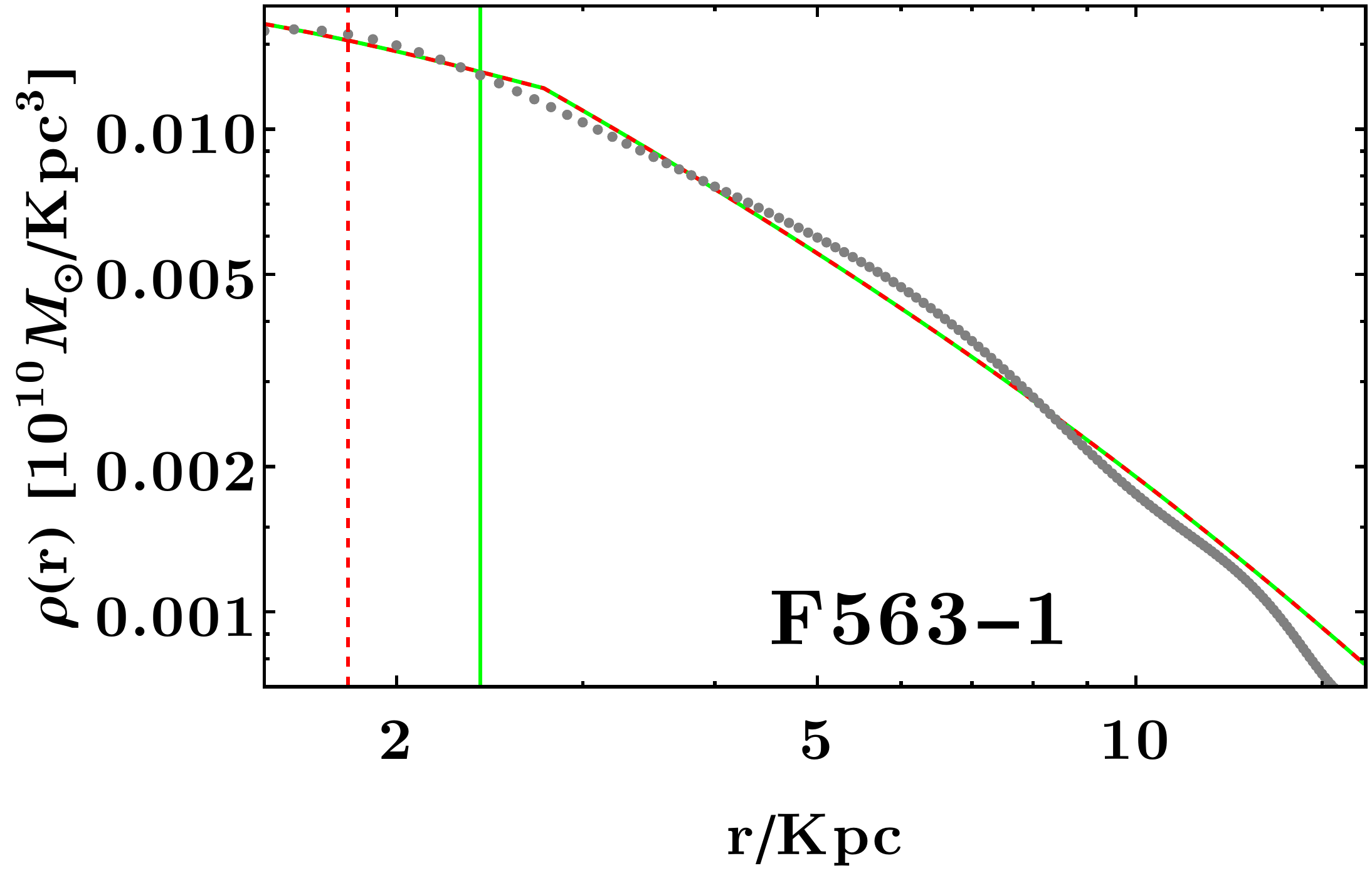}}
\subfloat[\label{sf:Fl2}]{\includegraphics[scale=0.147]{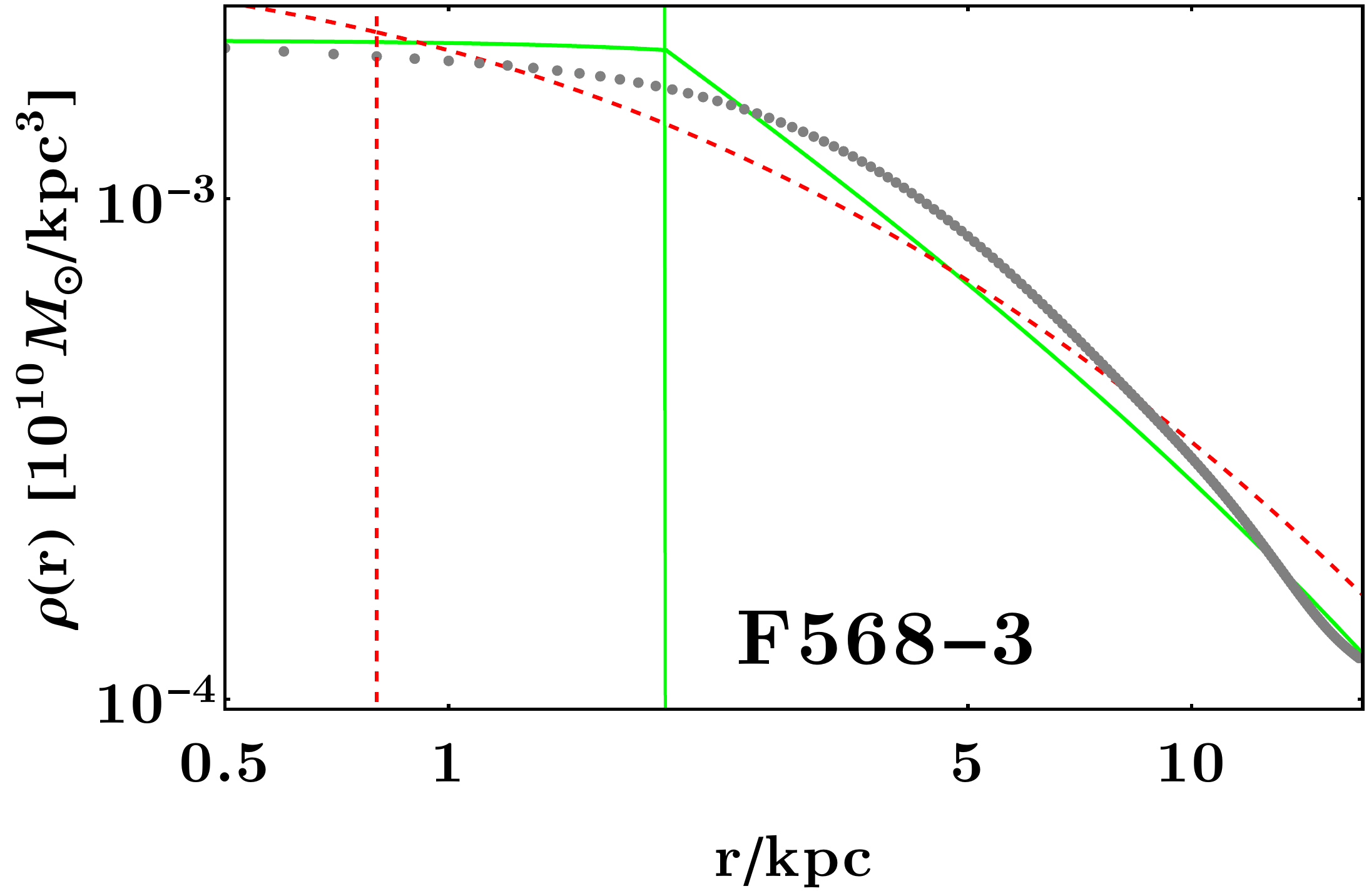}}
\subfloat[\label{sf:Fl3}]{\includegraphics[scale=0.147]{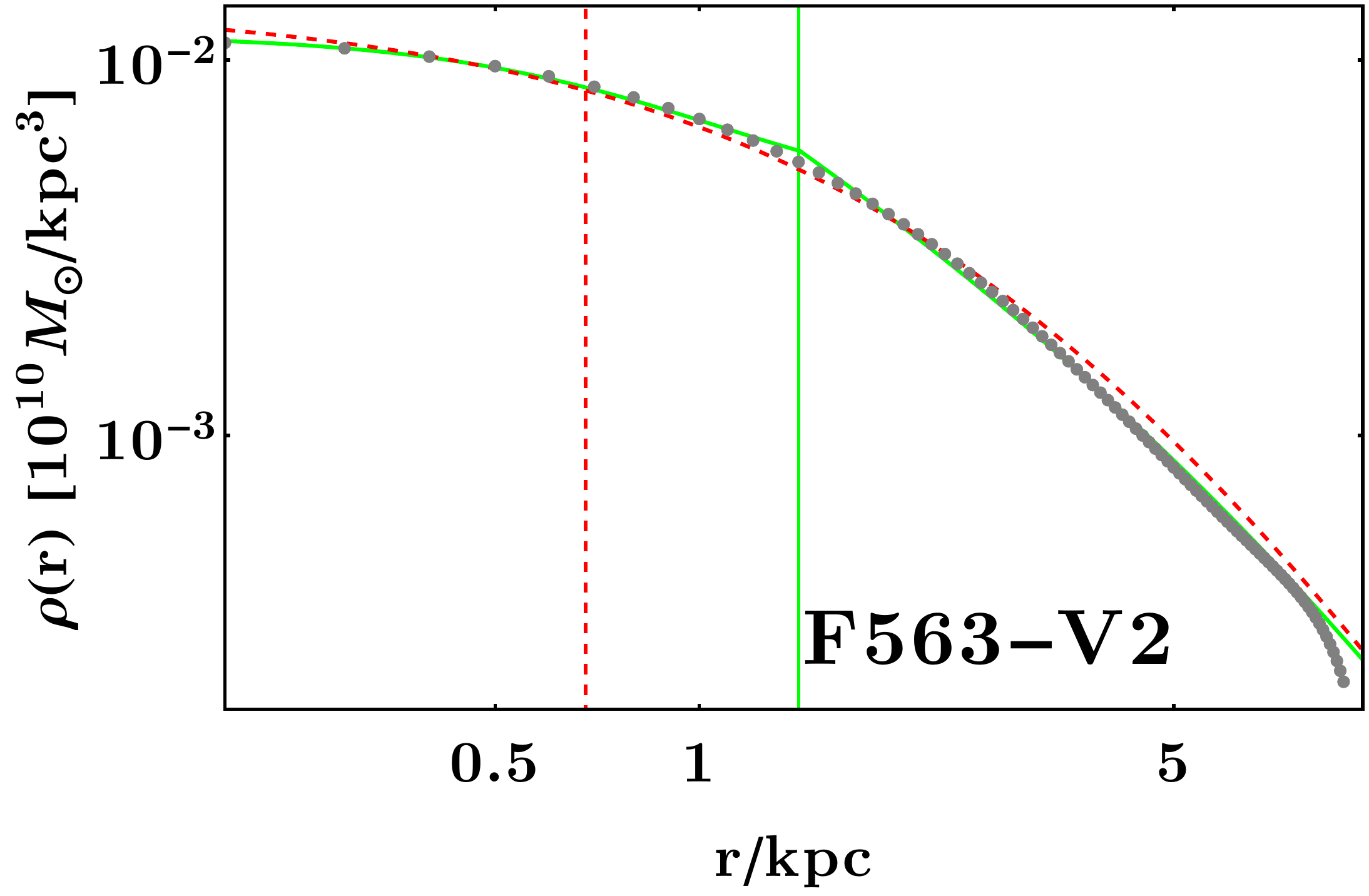}}
\linebreak
\subfloat[\label{sf:Fl4}]{\includegraphics[scale=0.147]{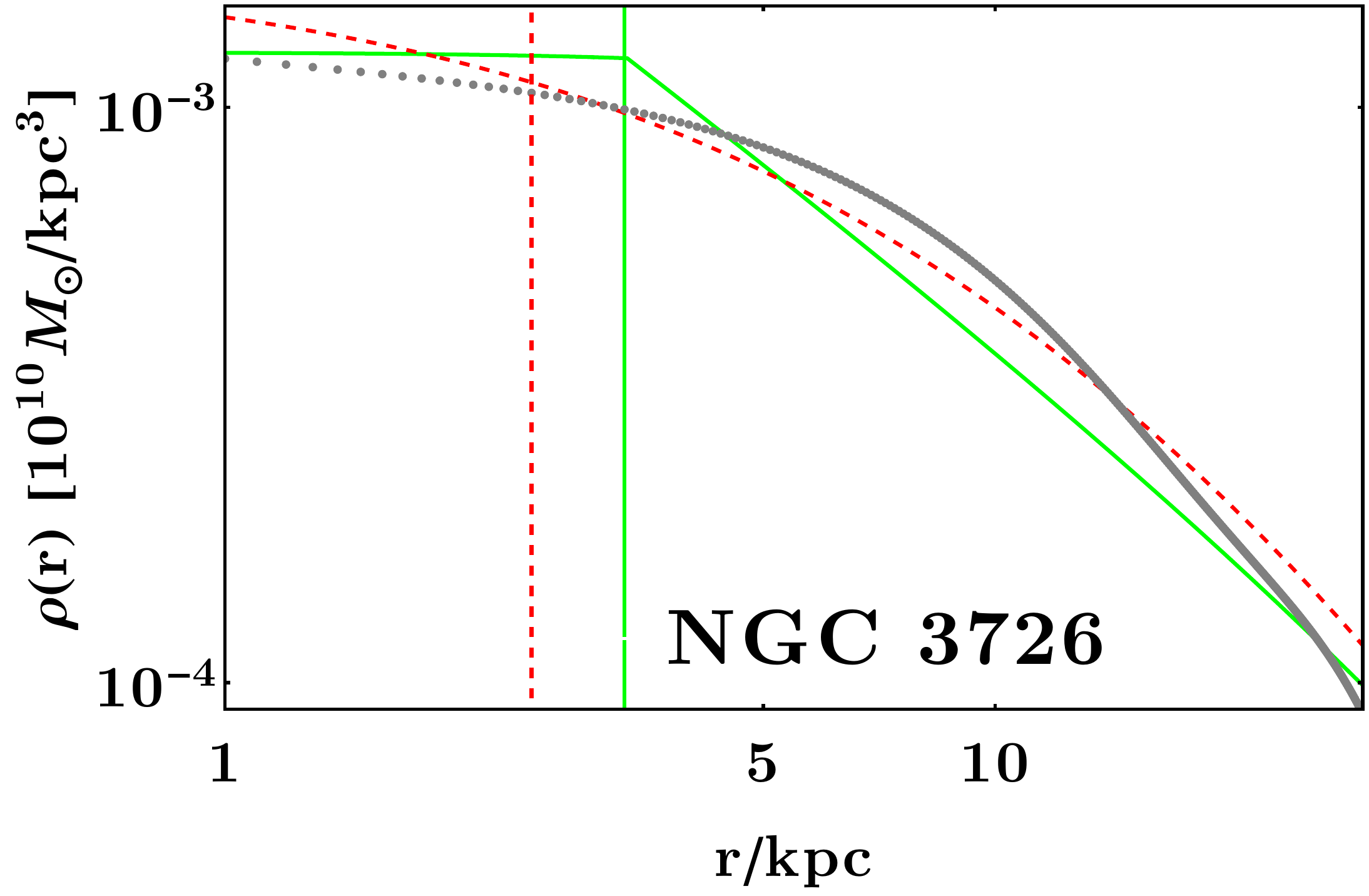}}
\subfloat[\label{sf:Fl5}]{\includegraphics[scale=0.147]{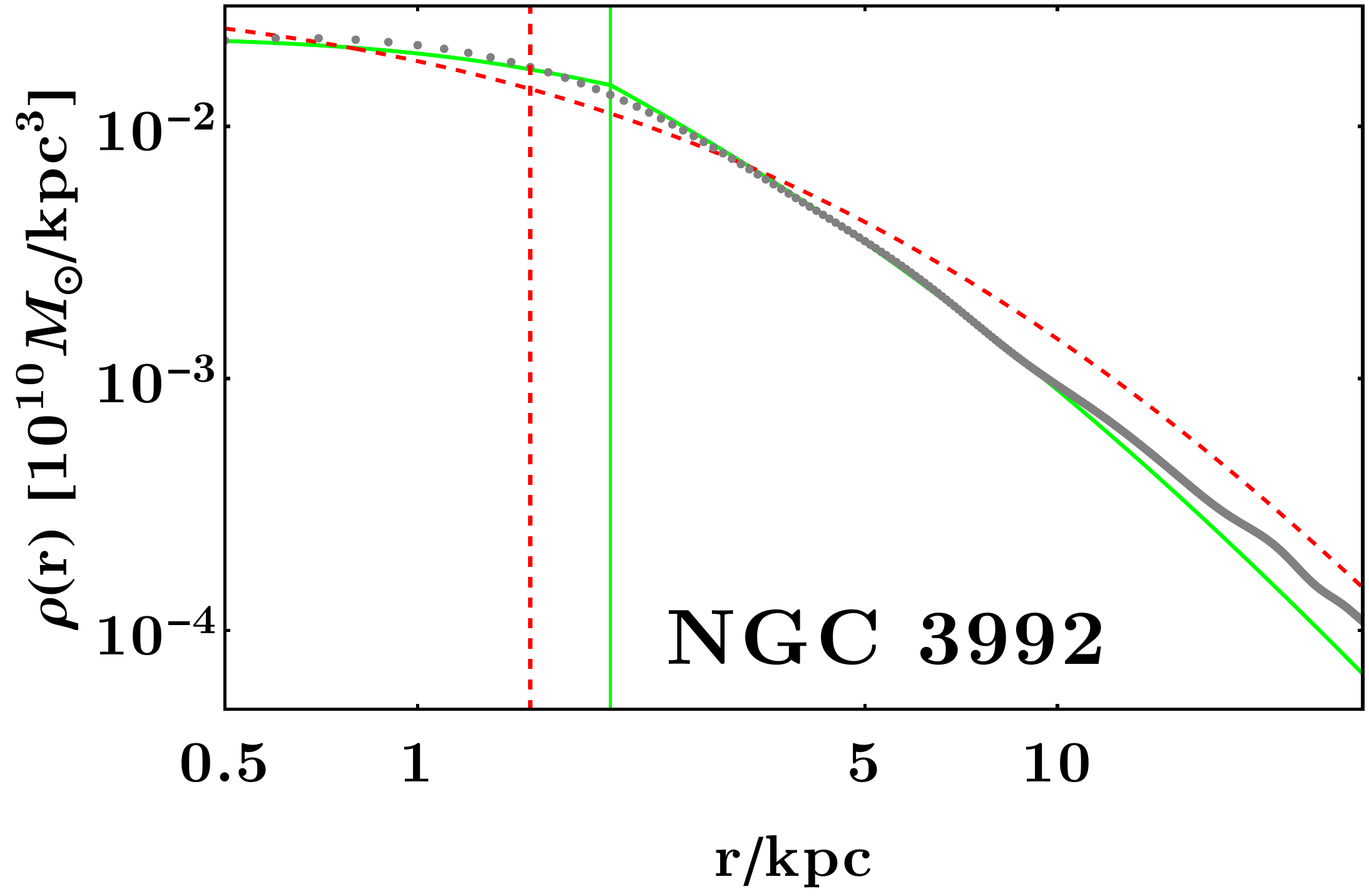}}
\subfloat[\label{sf:Fl6}]{\includegraphics[scale=0.147]{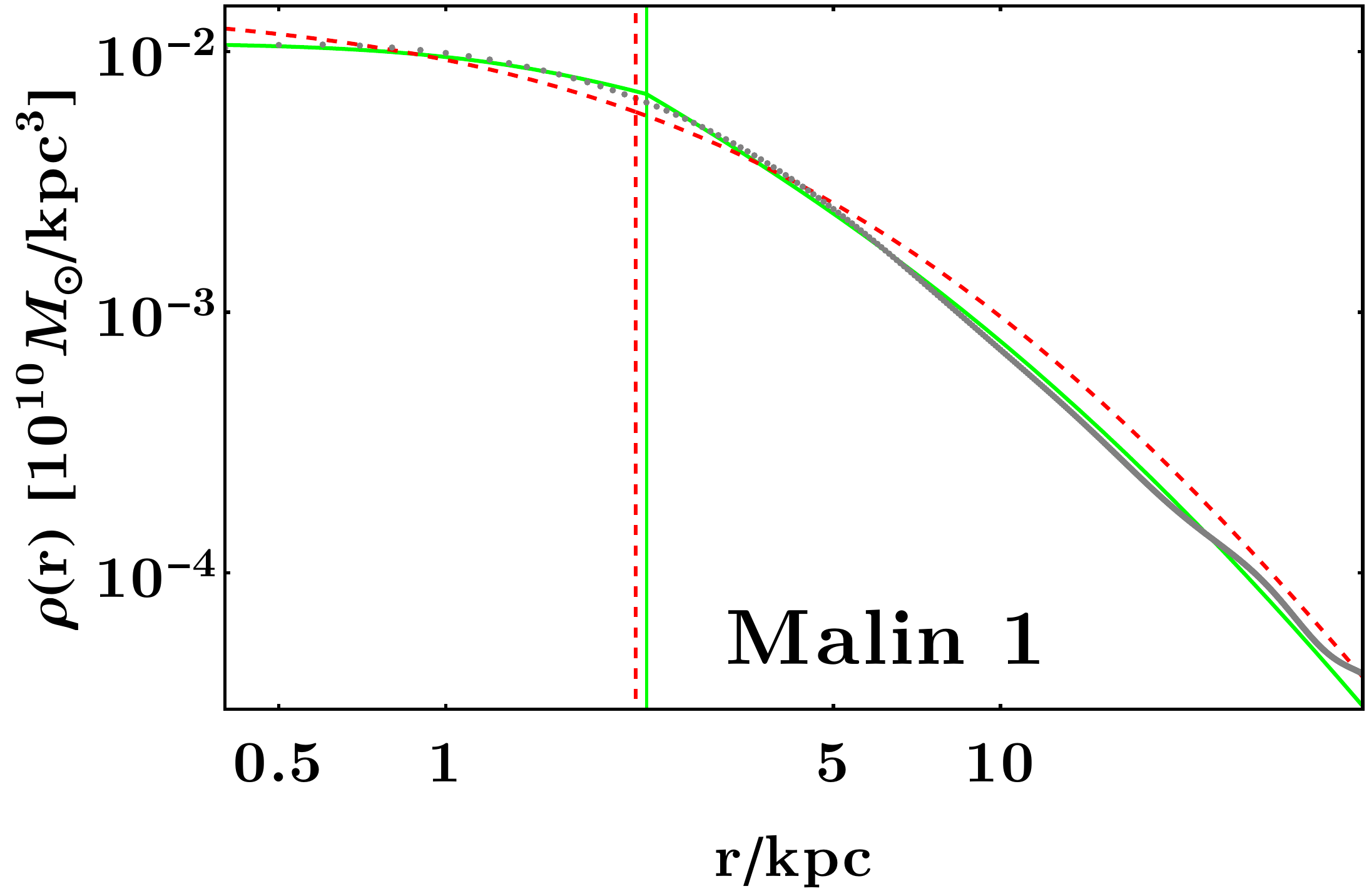}}
\caption{Model fits to the DM distributions inferred from LSB observations.}
\label{fig:appB2}
\end{center}
\end{figure*}

\begin{figure*}[h]
\begin{center}
\subfloat[\label{sf:Fc1}]{\includegraphics[scale=0.147]{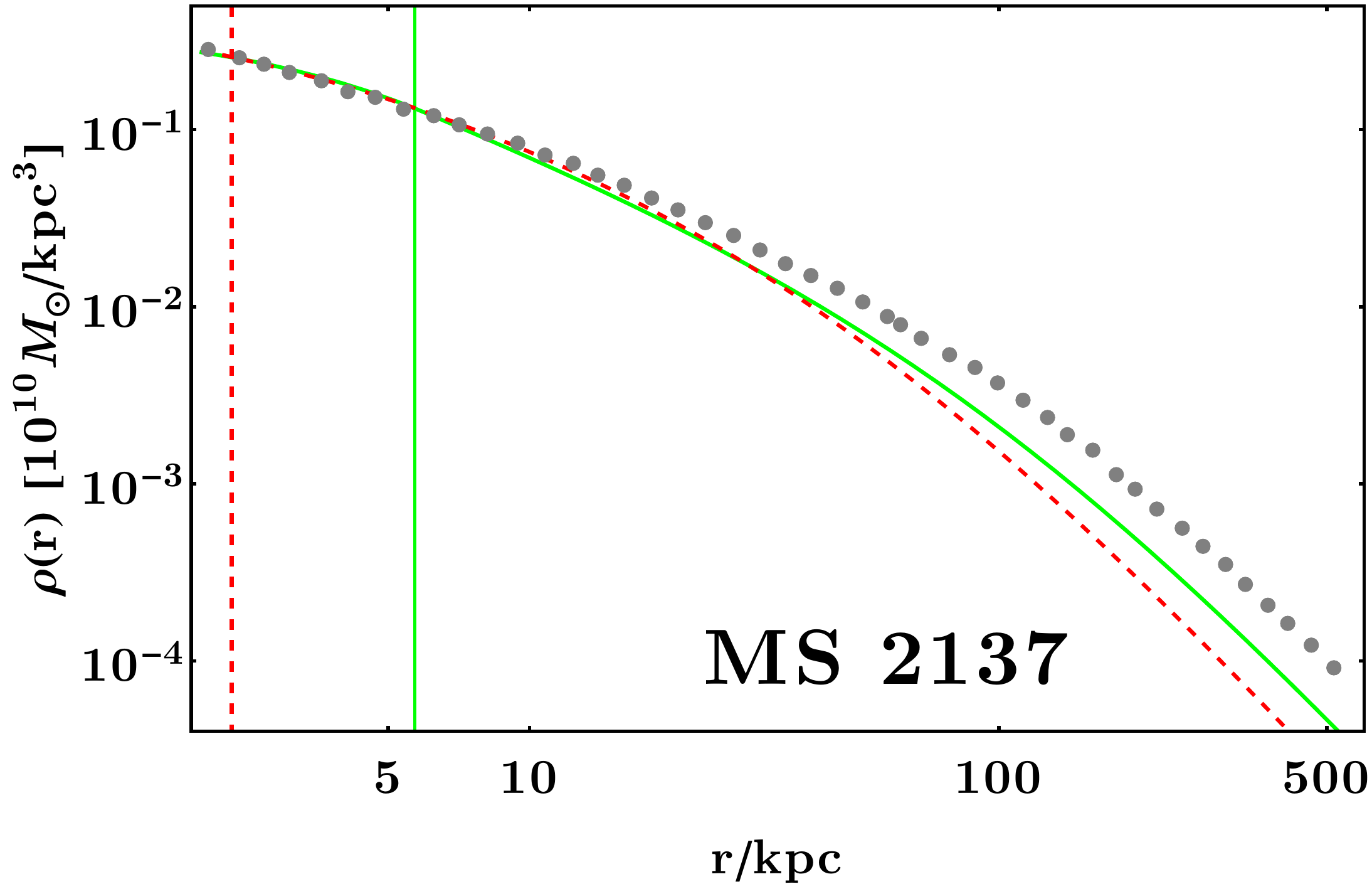}}
\subfloat[\label{sf:Fc2}]{\includegraphics[scale=0.147]{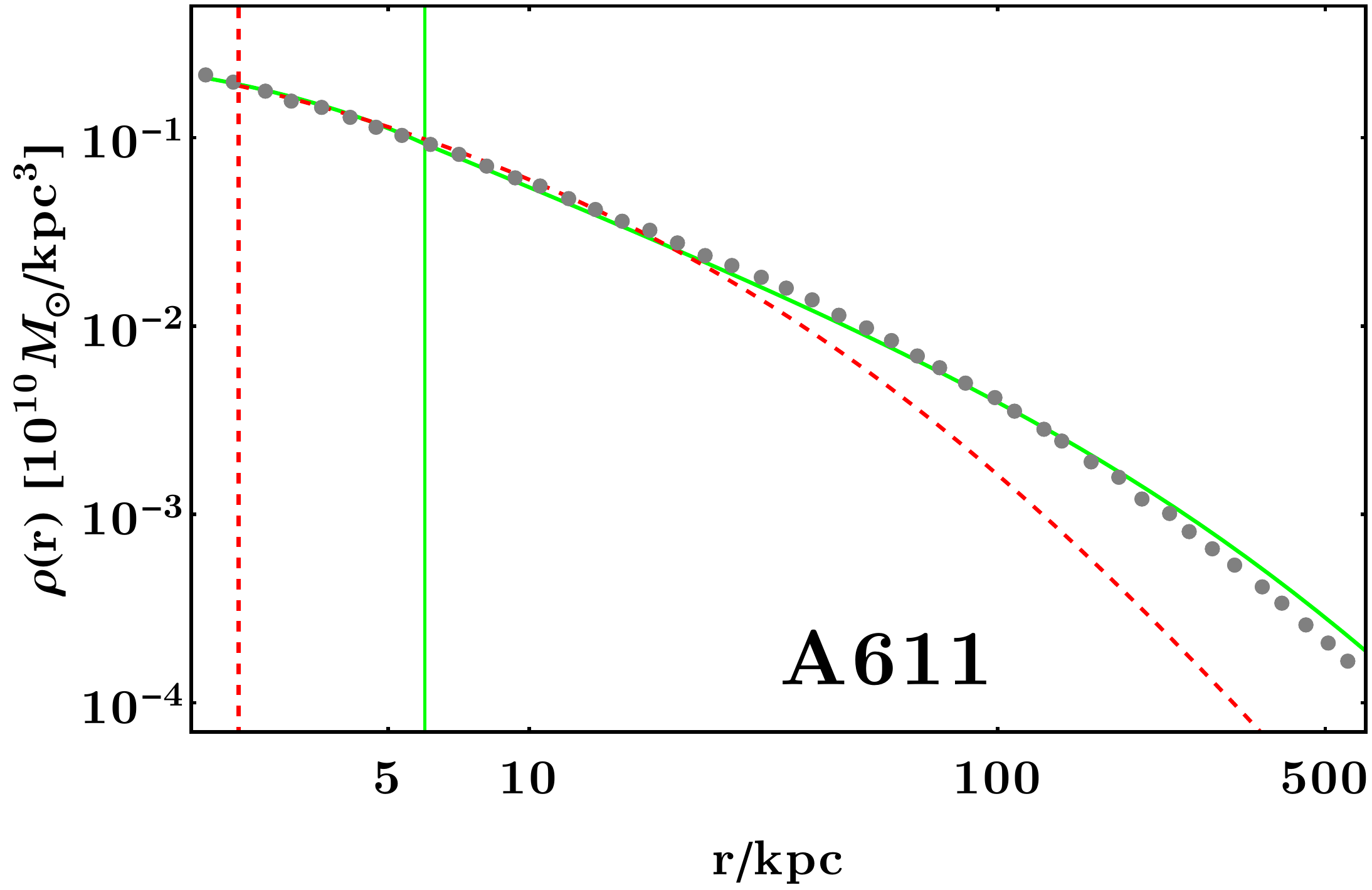}}
\subfloat[\label{sf:Fc3}]{\includegraphics[scale=0.147]{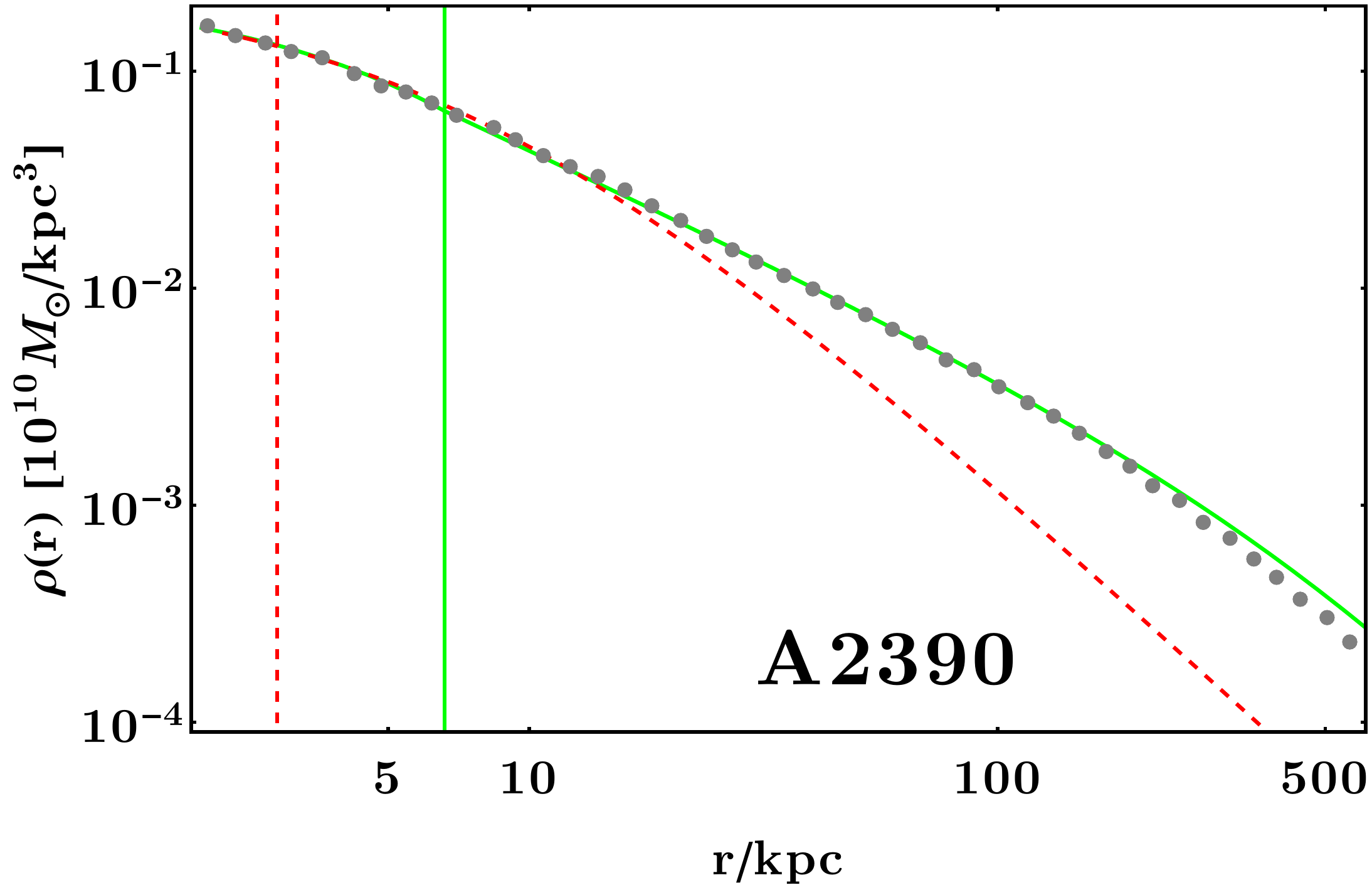}}
\caption{Model fits to the DM distributions inferred from cluster observations.}
\label{fig:appB3}
\end{center}
\end{figure*}

LSB and dwarf spheroidal galaxies are often considered as ideal targets for DM studies as their mass content of baryons are minimum. Observations indicate that many of these galaxies have a DM core rather than an expected cusp \cite{Simon:2004sr}. Here we present the individual fits to the inferred DM density distributions from the observational data for the dwarfs, LSBs and clusters in figure \ref{fig:appB1},\ref{fig:appB2} and \ref{fig:appB3} respectively, that have been discussed in section \ref{sec:OBScore}. The top,  middle  and bottom panels show the fits for the three dwarf spheroidals, six LSB galaxies and three clusters respectively. The green solid and red dashed curves represent the best fits to the inferred density distributions for the semi-analytic distribution and the cored NFW distribution respectively. The corresponding vertical lines denote the extent of the core radius. The DM distribution inferred from observations are shown by the gray points.

\FloatBarrier

\FloatBarrier
\section{Convergence tests}
\label{sec:appC}

\begin{figure*}[htbp]
\begin{center}
\subfloat[\label{sf:conver}]{\includegraphics[scale=0.3]{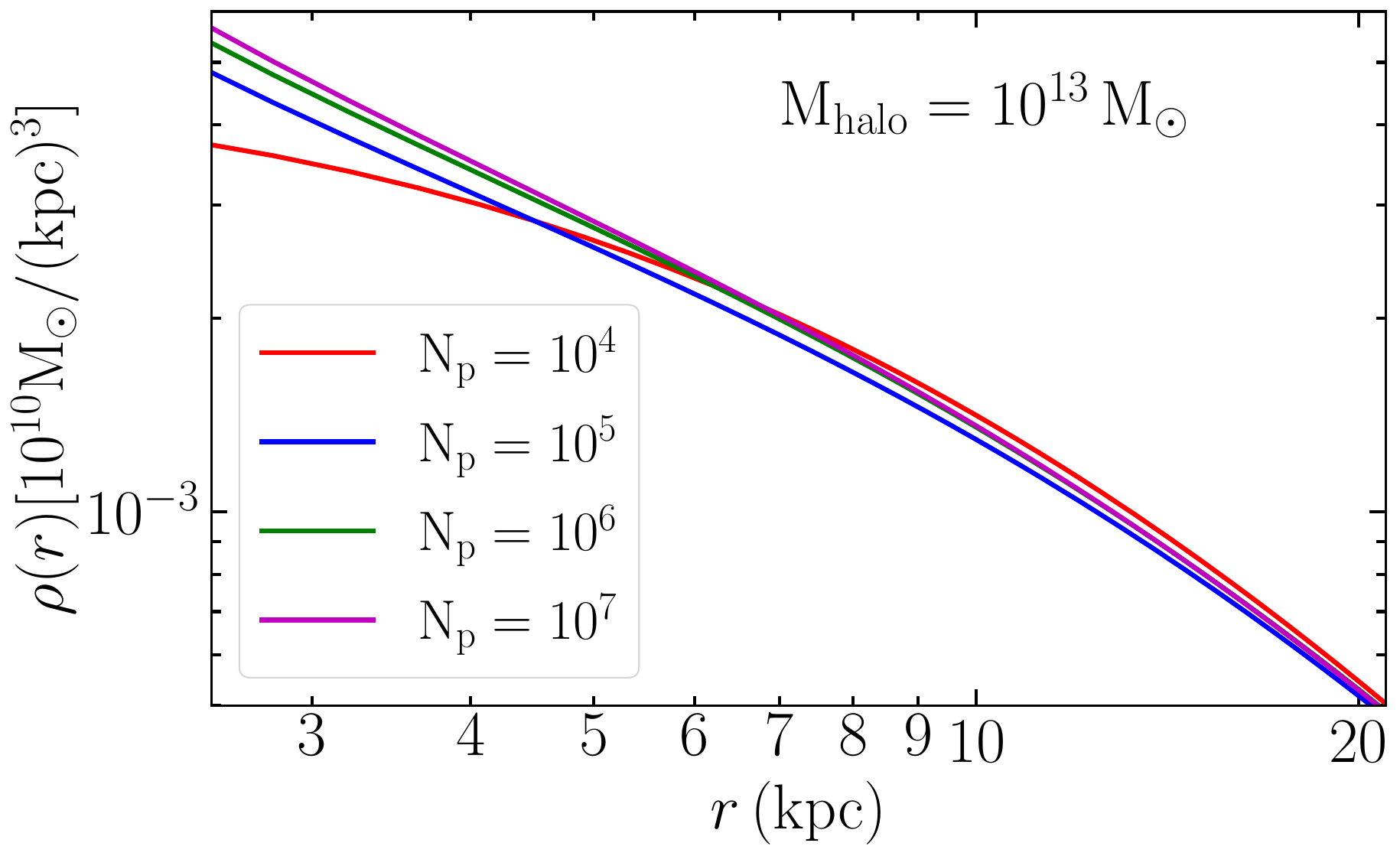}}
\caption{Variations in the central DM density for four different sets of particle numbers.}
\label{fig:appC}
\end{center}
\end{figure*}

Here we discuss the results of the convergence test for our simulations, with four different sets of particle numbers, $10^4$, $10^5$, $10^6$ and $10^7$ that have been shown in figure \ref{fig:appC} by the red, blue, green and purple curves respectively. We plot the DM density distribution of the halo as a function of the radius from its center for the CDM scenario with a halo mass of $10^{13}M_{\odot}$ in figure \ref{fig:appC}. We find a convergence in the central matter density as the particle content increases. The variation is seen to be less than $20\%$, with an order of magnitude change in particle numbers, which further decreases as the particle content of the simulations increase.
\bibliographystyle{JHEP}
\bibliography{interactivedm.bib}

\providecommand{\href}[2]{#2}\begingroup\raggedright\begin{thebibliography}{10}

\bibitem{1984Natur.311..517B}
G.~R. {Blumenthal}, S.~M. {Faber}, J.~R. {Primack} and M.~J. {Rees},
  \emph{{Formation of galaxies and large-scale structure with cold dark
  matter.}}, \href{http://dx.doi.org/10.1038/311517a0}{\emph{nat} {\bf 311}
  (Oct., 1984) 517--525}.

\bibitem{Lisanti:2016jxe}
M.~Lisanti, \emph{{Lectures on Dark Matter Physics}},  in \emph{{Theoretical
  Advanced Study Institute in Elementary Particle Physics}: {New Frontiers in
  Fields and Strings}}, pp.~399--446, 2017.
\newblock \href{https://arxiv.org/abs/1603.03797}{{\tt 1603.03797}}.
\newblock \href{http://dx.doi.org/10.1142/9789813149441_0007}{DOI}.

\bibitem{Pace:2019vrs}
F.~Pace, Z.~Sakr and I.~Tutusaus, \emph{{Spherical collapse in Generalized Dark
  Matter models}},
  \href{http://dx.doi.org/10.1103/PhysRevD.102.043512}{\emph{Phys. Rev. D} {\bf
  102} (2020) 043512}, [\href{https://arxiv.org/abs/1912.12250}{{\tt
  1912.12250}}].

\bibitem{DelPopolo:2002sz}
A.~Del~Popolo, \emph{{Dark matter, density perturbations and structure
  formation}},  \href{https://arxiv.org/abs/astro-ph/0209128}{{\tt
  astro-ph/0209128}}.

\bibitem{Bernardeau:2001qr}
F.~Bernardeau, S.~Colombi, E.~Gaztanaga and R.~Scoccimarro, \emph{{Large scale
  structure of the universe and cosmological perturbation theory}},
  \href{http://dx.doi.org/10.1016/S0370-1573(02)00135-7}{\emph{Phys. Rept.}
  {\bf 367} (2002) 1--248}, [\href{https://arxiv.org/abs/astro-ph/0112551}{{\tt
  astro-ph/0112551}}].

\bibitem{BoylanKolchin:2003sf}
M.~Boylan-Kolchin and C.-P. Ma, \emph{{Major mergers of galaxy haloes: Cuspy or
  cored inner density profile?}},
  \href{http://dx.doi.org/10.1111/j.1365-2966.2004.07585.x}{\emph{Mon. Not.
  Roy. Astron. Soc.} {\bf 349} (2004) 1117},
  [\href{https://arxiv.org/abs/astro-ph/0309243}{{\tt astro-ph/0309243}}].

\bibitem{BoylanKolchin:2011de}
M.~Boylan-Kolchin, J.~S. Bullock and M.~Kaplinghat, \emph{{Too big to fail? The
  puzzling darkness of massive Milky Way subhaloes}},
  \href{http://dx.doi.org/10.1111/j.1745-3933.2011.01074.x}{\emph{Mon. Not.
  Roy. Astron. Soc.} {\bf 415} (2011) L40},
  [\href{https://arxiv.org/abs/1103.0007}{{\tt 1103.0007}}].

\bibitem{10.1111/j.1365-2966.2012.20571.x}
A.~Pontzen and F.~Governato, \emph{{How supernova feedback turns dark matter
  cusps into cores}},
  \href{http://dx.doi.org/10.1111/j.1365-2966.2012.20571.x}{\emph{Monthly
  Notices of the Royal Astronomical Society} {\bf 421} (04, 2012) 3464--3471}.

\bibitem{Kim:2017iwr}
S.~Y. Kim, A.~H.~G. Peter and J.~R. Hargis, \emph{{Missing Satellites Problem:
  Completeness Corrections to the Number of Satellite Galaxies in the Milky Way
  are Consistent with Cold Dark Matter Predictions}},
  \href{http://dx.doi.org/10.1103/PhysRevLett.121.211302}{\emph{Phys. Rev.
  Lett.} {\bf 121} (2018) 211302},
  [\href{https://arxiv.org/abs/1711.06267}{{\tt 1711.06267}}].

\bibitem{10.1111/j.1365-2966.2010.16613.x}
A.~R. Duffy, J.~Schaye, S.~T. Kay, C.~D. Vecchia, R.~A. Battye and C.~M. Booth,
  \emph{{Impact of baryon physics on dark matter structures: a detailed
  simulation study of halo density profiles}},
  \href{http://dx.doi.org/10.1111/j.1365-2966.2010.16613.x}{\emph{Monthly
  Notices of the Royal Astronomical Society} {\bf 405} (07, 2010) 2161--2178}.

\bibitem{Sean:2017sea}
H.-B.~Y. Sean~Tulin, \emph{{Dark Matter Self-interactions and Small Scale
  Structure}},
  \href{http://dx.doi.org/10.1016/j.physrep.2017.11.004}{\emph{Physics Reports}
  {\bf 730} (2017) }, [\href{https://arxiv.org/abs/1705.02358}{{\tt
  1705.02358}}].

\bibitem{Kamada:2016euw}
A.~Kamada, M.~Kaplinghat, A.~B. Pace and H.-B. Yu, \emph{{How the
  Self-Interacting Dark Matter Model Explains the Diverse Galactic Rotation
  Curves}}, \href{http://dx.doi.org/10.1103/PhysRevLett.119.111102}{\emph{Phys.
  Rev. Lett.} {\bf 119} (2017) 111102},
  [\href{https://arxiv.org/abs/1611.02716}{{\tt 1611.02716}}].

\bibitem{Flores:1994gz}
R.~A. Flores and J.~R. Primack, \emph{{Observational and theoretical
  constraints on singular dark matter halos}},
  \href{http://dx.doi.org/10.1086/187350}{\emph{Astrophys. J. Lett.} {\bf 427}
  (1994) L1--4}, [\href{https://arxiv.org/abs/astro-ph/9402004}{{\tt
  astro-ph/9402004}}].

\bibitem{Burkert:1995yz}
A.~Burkert, \emph{{The Structure of dark matter halos in dwarf galaxies}},
  \href{http://dx.doi.org/10.1086/309560}{\emph{Astrophys. J. Lett.} {\bf 447}
  (1995) L25}, [\href{https://arxiv.org/abs/astro-ph/9504041}{{\tt
  astro-ph/9504041}}].

\bibitem{Spergel:1999mh}
D.~N. Spergel and P.~J. Steinhardt, \emph{{Observational evidence for
  self-interacting cold dark matter}},
  \href{http://dx.doi.org/10.1103/PhysRevLett.84.3760}{\emph{Phys. Rev. Lett.}
  {\bf 84} (2000) 3760--3763},
  [\href{https://arxiv.org/abs/astro-ph/9909386}{{\tt astro-ph/9909386}}].

\bibitem{Khlopov:2013ava}
M.~Khlopov, \emph{{Fundamental Particle Structure in the Cosmological Dark
  Matter}}, \href{http://dx.doi.org/10.1142/S0217751X13300421}{\emph{Int. J.
  Mod. Phys. A} {\bf 28} (2013) 1330042},
  [\href{https://arxiv.org/abs/1311.2468}{{\tt 1311.2468}}].

\bibitem{Huo:2019yhk}
R.~Huo, H.-B. Yu and Y.-M. Zhong, \emph{{The Structure of Dissipative Dark
  Matter Halos}},
  \href{http://dx.doi.org/10.1088/1475-7516/2020/06/051}{\emph{JCAP} {\bf 06}
  (2020) 051}, [\href{https://arxiv.org/abs/1912.06757}{{\tt 1912.06757}}].

\bibitem{Kamada:2020buc}
A.~Kamada, H.~J. Kim and T.~Kuwahara, \emph{{Maximally self-interacting dark
  matter: models and predictions}},
  \href{http://dx.doi.org/10.1007/JHEP12(2020)202}{\emph{JHEP} {\bf 20} (2020)
  202}, [\href{https://arxiv.org/abs/2007.15522}{{\tt 2007.15522}}].

\bibitem{Rocha:2012jg}
M.~Rocha, A.~H.~G. Peter, J.~S. Bullock, M.~Kaplinghat, S.~Garrison-Kimmel,
  J.~Onorbe et~al., \emph{{Cosmological Simulations with Self-Interacting Dark
  Matter I: Constant Density Cores and Substructure}},
  \href{http://dx.doi.org/10.1093/mnras/sts514}{\emph{Mon. Not. Roy. Astron.
  Soc.} {\bf 430} (2013) 81--104}, [\href{https://arxiv.org/abs/1208.3025}{{\tt
  1208.3025}}].

\bibitem{Elbert:2014bma}
O.~D. Elbert, J.~S. Bullock, S.~Garrison-Kimmel, M.~Rocha, J.~O\~norbe and
  A.~H.~G. Peter, \emph{{Core formation in dwarf haloes with self-interacting
  dark matter: no fine-tuning necessary}},
  \href{http://dx.doi.org/10.1093/mnras/stv1470}{\emph{Mon. Not. Roy. Astron.
  Soc.} {\bf 453} (2015) 29--37}, [\href{https://arxiv.org/abs/1412.1477}{{\tt
  1412.1477}}].

\bibitem{10.1093/mnrasl/sls053}
J.~Zavala, M.~Vogelsberger and M.~G. Walker, \emph{{Constraining
  self-interacting dark matter with the Milky Way’s dwarf spheroidals}},
  \href{http://dx.doi.org/10.1093/mnrasl/sls053}{\emph{Monthly Notices of the
  Royal Astronomical Society: Letters} {\bf 431} (02, 2013) L20--L24}.

\bibitem{Bondarenko:2017rfu}
K.~Bondarenko, A.~Boyarsky, T.~Bringmann and A.~Sokolenko, \emph{{Constraining
  self-interacting dark matter with scaling laws of observed halo surface
  densities}},
  \href{http://dx.doi.org/10.1088/1475-7516/2018/04/049}{\emph{JCAP} {\bf 04}
  (2018) 049}, [\href{https://arxiv.org/abs/1712.06602}{{\tt 1712.06602}}].

\bibitem{Correa_2021}
C.~A. Correa, \emph{Constraining velocity-dependent self-interacting dark
  matter with the milky way's dwarf spheroidal galaxies},
  \href{http://dx.doi.org/10.1093/mnras/stab506}{\emph{Monthly Notices of the
  Royal Astronomical Society} {\bf 503} (feb, 2021) 920--937}.

\bibitem{Randall:2007ph}
S.~W. Randall, M.~Markevitch, D.~Clowe, A.~H. Gonzalez and M.~Bradac,
  \emph{{Constraints on the Self-Interaction Cross-Section of Dark Matter from
  Numerical Simulations of the Merging Galaxy Cluster 1E 0657-56}},
  \href{http://dx.doi.org/10.1086/587859}{\emph{Astrophys. J.} {\bf 679} (2008)
  1173--1180}, [\href{https://arxiv.org/abs/0704.0261}{{\tt 0704.0261}}].

\bibitem{Peter:2012jh}
A.~H.~G. Peter, M.~Rocha, J.~S. Bullock and M.~Kaplinghat, \emph{{Cosmological
  Simulations with Self-Interacting Dark Matter II: Halo Shapes vs.
  Observations}}, \href{http://dx.doi.org/10.1093/mnras/sts535}{\emph{Mon. Not.
  Roy. Astron. Soc.} {\bf 430} (2013) 105},
  [\href{https://arxiv.org/abs/1208.3026}{{\tt 1208.3026}}].

\bibitem{Harvey:2015hha}
D.~Harvey, R.~Massey, T.~Kitching, A.~Taylor and E.~Tittley, \emph{{The
  non-gravitational interactions of dark matter in colliding galaxy clusters}},
  \href{http://dx.doi.org/10.1126/science.1261381}{\emph{Science} {\bf 347}
  (2015) 1462--1465}, [\href{https://arxiv.org/abs/1503.07675}{{\tt
  1503.07675}}].

\bibitem{Banerjee:2019bjp}
A.~Banerjee, S.~Adhikari, N.~Dalal, S.~More and A.~Kravtsov, \emph{{Signatures
  of Self-Interacting dark matter on cluster density profile and subhalo
  distributions}},
  \href{http://dx.doi.org/10.1088/1475-7516/2020/02/024}{\emph{JCAP} {\bf 02}
  (2020) 024}, [\href{https://arxiv.org/abs/1906.12026}{{\tt 1906.12026}}].

\bibitem{2010AdAst2010E...5D}
W.~J.~G. {de Blok}, \emph{{The Core-Cusp Problem}},
  \href{http://dx.doi.org/10.1155/2010/789293}{\emph{Advances in Astronomy}
  {\bf 2010} (Jan., 2010) 789293}, [\href{https://arxiv.org/abs/0910.3538}{{\tt
  0910.3538}}].

\bibitem{Salucci:2007tm}
P.~Salucci, A.~Lapi, C.~Tonini, G.~Gentile, I.~Yegorova and U.~Klein,
  \emph{{The Universal Rotation Curve of Spiral Galaxies. 2. The Dark Matter
  Distribution out to the Virial Radius}},
  \href{http://dx.doi.org/10.1111/j.1365-2966.2007.11696.x}{\emph{Mon. Not.
  Roy. Astron. Soc.} {\bf 378} (2007) 41--47},
  [\href{https://arxiv.org/abs/astro-ph/0703115}{{\tt astro-ph/0703115}}].

\bibitem{Nakano:1997sn}
T.~Nakano and J.~Makino, \emph{{On the origin of density cusps in elliptical
  galaxies}}, \href{http://dx.doi.org/10.1086/306547}{\emph{Astrophys. J.} {\bf
  510} (1999) 155}, [\href{https://arxiv.org/abs/astro-ph/9710135}{{\tt
  astro-ph/9710135}}].

\bibitem{Newman_2013n}
A.~B. Newman, T.~Treu, R.~S. Ellis and D.~J. Sand, \emph{The density profiles
  of massive, relaxed galaxy clusters. ii. separating luminous and dark matter
  in cluster cores},
  \href{http://dx.doi.org/10.1088/0004-637x/765/1/25}{\emph{The Astrophysical
  Journal} {\bf 765} (Feb, 2013) 25}.

\bibitem{Tollet_2016}
E.~Tollet, A.~V. Macci{\`{o} }, A.~A. Dutton, G.~S. Stinson, L.~Wang, C.~Penzo
  et~al., \emph{{NIHAO} {\textendash} {IV}: core creation and destruction in
  dark matter density profiles across cosmic time},
  \href{http://dx.doi.org/10.1093/mnras/stv2856}{\emph{Monthly Notices of the
  Royal Astronomical Society} {\bf 456} (jan, 2016) 3542--3552}.

\bibitem{Read:2018pft}
J.~I. Read, M.~G. Walker and P.~Steger, \emph{{The case for a cold dark matter
  cusp in Draco}}, \href{http://dx.doi.org/10.1093/mnras/sty2286}{\emph{Mon.
  Not. Roy. Astron. Soc.} {\bf 481} (2018) 860--877},
  [\href{https://arxiv.org/abs/1805.06934}{{\tt 1805.06934}}].

\bibitem{Read:2018fxs}
J.~I. Read, M.~G. Walker and P.~Steger, \emph{{Dark matter heats up in dwarf
  galaxies}}, \href{http://dx.doi.org/10.1093/mnras/sty3404}{\emph{Mon. Not.
  Roy. Astron. Soc.} {\bf 484} (2019) 1401--1420},
  [\href{https://arxiv.org/abs/1808.06634}{{\tt 1808.06634}}].

\bibitem{Oman:2017vkl}
K.~A. Oman, A.~Marasco, J.~F. Navarro, C.~S. Frenk, J.~Schaye and
  A.~Ben\'\i{}tez-Llambay, \emph{{Non-circular motions and the diversity of
  dwarf galaxy rotation curves}},
  \href{http://dx.doi.org/10.1093/mnras/sty2687}{\emph{Mon. Not. Roy. Astron.
  Soc.} {\bf 482} (2019) 821--847},
  [\href{https://arxiv.org/abs/1706.07478}{{\tt 1706.07478}}].

\bibitem{Kamada:2019wjo}
A.~Kamada and H.~J. Kim, \emph{{Escalating core formation with dark matter
  self-heating}},
  \href{http://dx.doi.org/10.1103/PhysRevD.102.043009}{\emph{Phys. Rev. D} {\bf
  102} (2020) 043009}, [\href{https://arxiv.org/abs/1911.09717}{{\tt
  1911.09717}}].

\bibitem{Kaplinghat:2015aga}
M.~Kaplinghat, S.~Tulin and H.-B. Yu, \emph{{Dark Matter Halos as Particle
  Colliders: Unified Solution to Small-Scale Structure Puzzles from Dwarfs to
  Clusters}},
  \href{http://dx.doi.org/10.1103/PhysRevLett.116.041302}{\emph{Phys. Rev.
  Lett.} {\bf 116} (2016) 041302},
  [\href{https://arxiv.org/abs/1508.03339}{{\tt 1508.03339}}].

\bibitem{Sokolenko:2018noz}
A.~Sokolenko, K.~Bondarenko, T.~Brinckmann, J.~Zavala, M.~Vogelsberger,
  T.~Bringmann et~al., \emph{{Towards an improved model of self-interacting
  dark matter haloes}},
  \href{http://dx.doi.org/10.1088/1475-7516/2018/12/038}{\emph{JCAP} {\bf 12}
  (2018) 038}, [\href{https://arxiv.org/abs/1806.11539}{{\tt 1806.11539}}].

\bibitem{Andrade:2020lqq}
K.~E. Andrade, J.~Fuson, S.~Gad-Nasr, D.~Kong, Q.~Minor, M.~G. Roberts et~al.,
  \emph{{A stringent upper limit on dark matter self-interaction cross-section
  from cluster strong lensing}},
  \href{http://dx.doi.org/10.1093/mnras/stab3241}{\emph{Mon. Not. Roy. Astron.
  Soc.} {\bf 510} (2021) 54--81}, [\href{https://arxiv.org/abs/2012.06611}{{\tt
  2012.06611}}].

\bibitem{Sagunski:2020spe}
L.~Sagunski, S.~Gad-Nasr, B.~Colquhoun, A.~Robertson and S.~Tulin,
  \emph{{Velocity-dependent Self-interacting Dark Matter from Groups and
  Clusters of Galaxies}},
  \href{http://dx.doi.org/10.1088/1475-7516/2021/01/024}{\emph{JCAP} {\bf 01}
  (2021) 024}, [\href{https://arxiv.org/abs/2006.12515}{{\tt 2006.12515}}].

\bibitem{Springel:2000yr}
V.~Springel, N.~Yoshida and S.~D.~M. White, \emph{{GADGET: A Code for
  collisionless and gasdynamical cosmological simulations}},
  \href{http://dx.doi.org/10.1016/S1384-1076(01)00042-2}{\emph{New Astron.}
  {\bf 6} (2001) 79}, [\href{https://arxiv.org/abs/astro-ph/0003162}{{\tt
  astro-ph/0003162}}].

\bibitem{Springel:2005mi}
V.~Springel, \emph{{The Cosmological simulation code GADGET-2}},
  \href{http://dx.doi.org/10.1111/j.1365-2966.2005.09655.x}{\emph{Mon. Not.
  Roy. Astron. Soc.} {\bf 364} (2005) 1105--1134},
  [\href{https://arxiv.org/abs/astro-ph/0505010}{{\tt astro-ph/0505010}}].

\bibitem{Jun:2011jun}
P.~R.~S. Jun~Koda, \emph{{Gravothermal collapse of isolated self-interacting
  dark matter haloes: N-body simulation versus the fluid model}},
  \href{http://dx.doi.org/10.1111/j.1365-2966.2011.18684.x}{\emph{Mon. Not. R.
  Astron. Soc} (2011) }, [\href{https://arxiv.org/abs/1101.3097}{{\tt
  1101.3097}}].

\bibitem{Zemp:2007nt}
M.~Zemp, B.~Moore, J.~Stadel, C.~M. Carollo and P.~Madau, \emph{{Multi-Mass
  Spherical Structure Models for N-body Simulations}},
  \href{http://dx.doi.org/10.1111/j.1365-2966.2008.13126.x}{\emph{Mon. Not.
  Roy. Astron. Soc.} {\bf 386} (2008) 1543},
  [\href{https://arxiv.org/abs/0710.3189}{{\tt 0710.3189}}].

\bibitem{Colin:2002nk}
P.~Colin, V.~Avila-Reese, O.~Valenzuela and C.~Firmani, \emph{{Structure and
  subhalo population of halos in a selfinteracting dark matter cosmology}},
  \href{http://dx.doi.org/10.1086/344259}{\emph{Astrophys. J.} {\bf 581} (2002)
  777--793}, [\href{https://arxiv.org/abs/astro-ph/0205322}{{\tt
  astro-ph/0205322}}].

\bibitem{Robertson:2016qef}
A.~Robertson, R.~Massey and V.~Eke, \emph{{Cosmic particle colliders:
  simulations of self-interacting dark matter with anisotropic scattering}},
  \href{http://dx.doi.org/10.1093/mnras/stx463}{\emph{Mon. Not. Roy. Astron.
  Soc.} {\bf 467} (2017) 4719--4730},
  [\href{https://arxiv.org/abs/1612.03906}{{\tt 1612.03906}}].

\bibitem{Drakos:2017gfy}
N.~E. Drakos, J.~E. Taylor and A.~J. Benson, \emph{{The phase-space structure
  of tidally stripped haloes}},
  \href{http://dx.doi.org/10.1093/mnras/stx652}{\emph{Mon. Not. Roy. Astron.
  Soc.} {\bf 468} (2017) 2345--2358},
  [\href{https://arxiv.org/abs/1703.07836}{{\tt 1703.07836}}].

\bibitem{2017MNRAS.470.4941H}
J.~{Herpich}, G.~S. {Stinson}, H.~W. {Rix}, M.~{Martig} and A.~A. {Dutton},
  \emph{{How to bend galaxy disc profiles - II. Stars surfing the bar in
  Type-III discs}}, \href{http://dx.doi.org/10.1093/mnras/stx1511}{\emph{mnras}
  {\bf 470} (Oct., 2017) 4941--4955},
  [\href{https://arxiv.org/abs/1511.04442}{{\tt 1511.04442}}].

\bibitem{Fischer:2020uxh}
M.~S. Fischer, M.~Br\"uggen, K.~Schmidt-Hoberg, K.~Dolag, F.~Kahlhoefer,
  A.~Ragagnin et~al., \emph{{N-body simulations of dark matter with frequent
  self-interactions}},
  \href{http://dx.doi.org/10.1093/mnras/stab1198}{\emph{Mon. Not. Roy. Astron.
  Soc.} {\bf 505} (2021) 851--868},
  [\href{https://arxiv.org/abs/2012.10277}{{\tt 2012.10277}}].

\bibitem{Burger:2018sqp}
J.~D. Burger and J.~Zavala, \emph{{The nature of core formation in dark matter
  haloes: adiabatic or impulsive?}},
  \href{http://dx.doi.org/10.1093/mnras/stz496}{\emph{Mon. Not. Roy. Astron.
  Soc.} {\bf 485} (2019) 1008}, [\href{https://arxiv.org/abs/1810.10024}{{\tt
  1810.10024}}].

\bibitem{Ade:2015xua}
{\scshape Planck} collaboration, P.~A.~R. Ade et~al., \emph{{Planck 2015
  results. XIII. Cosmological parameters}},
  \href{http://dx.doi.org/10.1051/0004-6361/201525830}{\emph{Astron.
  Astrophys.} {\bf 594} (2016) A13},
  [\href{https://arxiv.org/abs/1502.01589}{{\tt 1502.01589}}].

\bibitem{Miralda-Escude:2000tvu}
J.~Miralda-Escude, \emph{{A test of the collisional dark matter hypothesis from
  cluster lensing}}, \href{http://dx.doi.org/10.1086/324138}{\emph{Astrophys.
  J.} {\bf 564} (2002) 60}, [\href{https://arxiv.org/abs/astro-ph/0002050}{{\tt
  astro-ph/0002050}}].

\bibitem{Feng:2009hw}
J.~L. Feng, M.~Kaplinghat and H.-B. Yu, \emph{{Halo Shape and Relic Density
  Exclusions of Sommerfeld-Enhanced Dark Matter Explanations of Cosmic Ray
  Excesses}},
  \href{http://dx.doi.org/10.1103/PhysRevLett.104.151301}{\emph{Phys. Rev.
  Lett.} {\bf 104} (2010) 151301}, [\href{https://arxiv.org/abs/0911.0422}{{\tt
  0911.0422}}].

\bibitem{Burkert:2000di}
A.~Burkert, \emph{{The Structure and evolution of weakly selfinteracting cold
  dark matter halos}}, \href{http://dx.doi.org/10.1086/312674}{\emph{Astrophys.
  J. Lett.} {\bf 534} (2000) L143--L146},
  [\href{https://arxiv.org/abs/astro-ph/0002409}{{\tt astro-ph/0002409}}].

\bibitem{Dave:2000ar}
R.~Dave, D.~N. Spergel, P.~J. Steinhardt and B.~D. Wandelt, \emph{{Halo
  properties in cosmological simulations of selfinteracting cold dark matter}},
  \href{http://dx.doi.org/10.1086/318417}{\emph{Astrophys. J.} {\bf 547} (2001)
  574--589}, [\href{https://arxiv.org/abs/astro-ph/0006218}{{\tt
  astro-ph/0006218}}].

\bibitem{Kaplinghat:2013yxa}
M.~Kaplinghat, S.~Tulin and H.-B. Yu, \emph{{Direct Detection Portals for
  Self-interacting Dark Matter}},
  \href{http://dx.doi.org/10.1103/PhysRevD.89.035009}{\emph{Phys. Rev. D} {\bf
  89} (2014) 035009}, [\href{https://arxiv.org/abs/1310.7945}{{\tt
  1310.7945}}].

\bibitem{Kaplinghat:2013xca}
M.~Kaplinghat, R.~E. Keeley, T.~Linden and H.-B. Yu, \emph{{Tying Dark Matter
  to Baryons with Self-interactions}},
  \href{http://dx.doi.org/10.1103/PhysRevLett.113.021302}{\emph{Phys. Rev.
  Lett.} {\bf 113} (2014) 021302}, [\href{https://arxiv.org/abs/1311.6524}{{\tt
  1311.6524}}].

\bibitem{Robertson:2020pxj}
A.~Robertson, R.~Massey, V.~Eke, J.~Schaye and T.~Theuns, \emph{{The surprising
  accuracy of isothermal Jeans modelling of self-interacting dark matter
  density profiles}},
  \href{http://dx.doi.org/10.1093/mnras/staa3954}{\emph{Mon. Not. Roy. Astron.
  Soc.} {\bf 501} (2021) 4610--4634},
  [\href{https://arxiv.org/abs/2009.07844}{{\tt 2009.07844}}].

\bibitem{Navarro:1995iw}
J.~F. Navarro, C.~S. Frenk and S.~D.~M. White, \emph{{The Structure of cold
  dark matter halos}}, \href{http://dx.doi.org/10.1086/177173}{\emph{Astrophys.
  J.} {\bf 462} (1996) 563--575},
  [\href{https://arxiv.org/abs/astro-ph/9508025}{{\tt astro-ph/9508025}}].

\bibitem{Valli:2017ktb}
M.~Valli and H.-B. Yu, \emph{{Dark matter self-interactions from the internal
  dynamics of dwarf spheroidals}},
  \href{http://dx.doi.org/10.1038/s41550-018-0560-7}{\emph{Nature Astron.} {\bf
  2} (2018) 907--912}, [\href{https://arxiv.org/abs/1711.03502}{{\tt
  1711.03502}}].

\bibitem{Lokas:2000mu}
E.~L. Lokas and G.~A. Mamon, \emph{{Properties of spherical galaxies and
  clusters with an nfw density profile}},
  \href{http://dx.doi.org/10.1046/j.1365-8711.2001.04007.x}{\emph{Mon. Not.
  Roy. Astron. Soc.} {\bf 321} (2001) 155},
  [\href{https://arxiv.org/abs/astro-ph/0002395}{{\tt astro-ph/0002395}}].

\bibitem{Lu:2005tu}
Y.~Lu, H.~J. Mo, N.~Katz and M.~D. Weinberg, \emph{{On the origin of cold dark
  matter halo density profiles}},
  \href{http://dx.doi.org/10.1111/j.1365-2966.2006.10270.x}{\emph{Mon. Not.
  Roy. Astron. Soc.} {\bf 368} (2006) 1931--1940},
  [\href{https://arxiv.org/abs/astro-ph/0508624}{{\tt astro-ph/0508624}}].

\bibitem{10.1093/mnras/stz1698}
H.-N. Lin and X.~Li, \emph{{The dark matter profiles in the Milky Way}},
  \href{http://dx.doi.org/10.1093/mnras/stz1698}{\emph{Monthly Notices of the
  Royal Astronomical Society} {\bf 487} (06, 2019) 5679--5684}.

\bibitem{Newman_2013}
A.~B. Newman, T.~Treu, R.~S. Ellis, D.~J. Sand, C.~Nipoti, J.~Richard et~al.,
  \emph{The density profiles of massive, relaxed galaxy clusters. i. the total
  density over three decades in radius},
  \href{http://dx.doi.org/10.1088/0004-637x/765/1/24}{\emph{The Astrophysical
  Journal} {\bf 765} (Feb, 2013) 24}.

\bibitem{Read:2015sta}
J.~I. Read, O.~Agertz and M.~L.~M. Collins, \emph{{Dark matter cores all the
  way down}}, \href{http://dx.doi.org/10.1093/mnras/stw713}{\emph{Mon. Not.
  Roy. Astron. Soc.} {\bf 459} (2016) 2573--2590},
  [\href{https://arxiv.org/abs/1508.04143}{{\tt 1508.04143}}].

\bibitem{Li:2020iib}
P.~Li, F.~Lelli, S.~McGaugh and J.~Schombert, \emph{{A comprehensive catalog of
  dark matter halo models for SPARC galaxies}},
  \href{http://dx.doi.org/10.3847/1538-4365/ab700e}{\emph{Astrophys. J. Suppl.}
  {\bf 247} (2020) 31}, [\href{https://arxiv.org/abs/2001.10538}{{\tt
  2001.10538}}].

\bibitem{1965TrAlm...5...87E}
J.~{Einasto}, \emph{{On the Construction of a Composite Model for the Galaxy
  and on the Determination of the System of Galactic Parameters}}, {\emph{Trudy
  Astrofizicheskogo Instituta Alma-Ata} {\bf 5} (Jan., 1965) 87--100}.

\bibitem{10.1093/mnras/249.3.523}
K.~G. Begeman, A.~H. Broeils and R.~H. Sanders, \emph{{Extended rotation curves
  of spiral galaxies: dark haloes and modified dynamics}},
  \href{http://dx.doi.org/10.1093/mnras/249.3.523}{\emph{Monthly Notices of the
  Royal Astronomical Society} {\bf 249} (04, 1991) 523--537}.

\bibitem{deBlok:2008wp}
W.~J.~G. de~Blok, F.~Walter, E.~Brinks, C.~Trachternach, S.-H. Oh and R.~C.
  Kennicutt, Jr., \emph{{High-Resolution Rotation Curves and Galaxy Mass Models
  from THINGS}},
  \href{http://dx.doi.org/10.1088/0004-6256/136/6/2648}{\emph{Astron. J.} {\bf
  136} (2008) 2648--2719}, [\href{https://arxiv.org/abs/0810.2100}{{\tt
  0810.2100}}].

\bibitem{Walter:2008wy}
F.~Walter, E.~Brinks, W.~J.~G. de~Blok, F.~Bigiel, R.~C. Kennicutt, Jr., M.~D.
  Thornley et~al., \emph{{THINGS: The HI Nearby Galaxy Survey}},
  \href{http://dx.doi.org/10.1088/0004-6256/136/6/2563}{\emph{Astron. J.} {\bf
  136} (1985) 2563--2647}, [\href{https://arxiv.org/abs/0810.2125}{{\tt
  0810.2125}}].

\bibitem{Kennicutt:2003dc}
R.~C. Kennicutt, Jr. et~al., \emph{{SINGS: The SIRTF Nearby Galaxies Survey}},
  \href{http://dx.doi.org/10.1086/376941}{\emph{Publ. Astron. Soc. Pac.} {\bf
  115} (2003) 928--952}, [\href{https://arxiv.org/abs/astro-ph/0305437}{{\tt
  astro-ph/0305437}}].

\bibitem{Oh_2011}
S.-H. Oh, C.~Brook, F.~Governato, E.~Brinks, L.~Mayer, W.~J.~G. de~Blok et~al.,
  \emph{The central slope of dark matter cores in dwarf galaxies: Simulations
  versus things}, \href{http://dx.doi.org/10.1088/0004-6256/142/1/24}{\emph{The
  Astronomical Journal} {\bf 142} (Jun, 2011) 24}.

\bibitem{KuziodeNaray:2006wh}
R.~Kuzio~de Naray, S.~S. McGaugh, W.~J.~G. de~Blok and A.~Bosma, \emph{{High
  Resolution Optical Velocity Fields of 11 Low Surface Brightness Galaxies}},
  \href{http://dx.doi.org/10.1086/505345}{\emph{Astrophys. J. Suppl.} {\bf 165}
  (2006) 461--479}, [\href{https://arxiv.org/abs/astro-ph/0604576}{{\tt
  astro-ph/0604576}}].

\bibitem{KuziodeNaray:2007qi}
R.~Kuzio~de Naray, S.~S. McGaugh and W.~J.~G. de~Blok, \emph{{Mass Models for
  Low Surface Brightness Galaxies with High Resolution Optical Velocity
  Fields}}, \href{http://dx.doi.org/10.1086/527543}{\emph{Astrophys. J.} {\bf
  676} (2008) 920--943}, [\href{https://arxiv.org/abs/0712.0860}{{\tt
  0712.0860}}].

\bibitem{article}
A.~Gusev, A.~Zasov, S.~Kaisin and D.~Bizyaev, \emph{Bvri surface photometry of
  the galaxy ngc 3726},
  \href{http://dx.doi.org/10.1134/1.1508062}{\emph{Astronomy Reports} {\bf 46}
  (09, 2002) 704--711}.

\bibitem{Craciun:2020twu}
M.~Cr\u{a}ciun and T.~Harko, \emph{{Testing Bose\textendash{}Einstein
  condensate dark matter models with the SPARC galactic rotation curves data}},
  \href{http://dx.doi.org/10.1140/epjc/s10052-020-8272-4}{\emph{Eur. Phys. J.
  C} {\bf 80} (2020) 735}, [\href{https://arxiv.org/abs/2007.12222}{{\tt
  2007.12222}}].

\bibitem{Bottema:2002yz}
R.~Bottema and M.~A.~W. Verheijen, \emph{{Dark and luminous matter in the ngc
  3992 group of galaxies, I. the large barred spiral ngc 3992}},
  \href{http://dx.doi.org/10.1051/0004-6361:20020539}{\emph{Astron. Astrophys.}
  {\bf 388} (2002) 793}, [\href{https://arxiv.org/abs/astro-ph/0204335}{{\tt
  astro-ph/0204335}}].

\bibitem{2010}
F.~Lelli, F.~Fraternali and R.~Sancisi, \emph{Structure and dynamics of giant
  low surface brightness galaxies},
  \href{http://dx.doi.org/10.1051/0004-6361/200913808}{\emph{Astronomy and
  Astrophysics} {\bf 516} (Jun, 2010) A11}.

\bibitem{2020}
Junais, S.~Boissier, B.~Epinat, P.~Amram, B.~F. Madore, A.~Boselli et~al.,
  \emph{First spectroscopic study of ionised gas emission lines in the extreme
  low surface brightness galaxy malin 1},
  \href{http://dx.doi.org/10.1051/0004-6361/201937330}{\emph{Astronomy
  Astrophysics} {\bf 637} (May, 2020) A21}.

\bibitem{Oman:2015xda}
K.~A. Oman et~al., \emph{{The unexpected diversity of dwarf galaxy rotation
  curves}}, \href{http://dx.doi.org/10.1093/mnras/stv1504}{\emph{Mon. Not. Roy.
  Astron. Soc.} {\bf 452} (2015) 3650--3665},
  [\href{https://arxiv.org/abs/1504.01437}{{\tt 1504.01437}}].

\bibitem{Virtanen:2019joe}
P.~Virtanen et~al., \emph{{SciPy 1.0--Fundamental Algorithms for Scientific
  Computing in Python}},
  \href{http://dx.doi.org/10.1038/s41592-019-0686-2}{\emph{Nature Meth.} {\bf
  17} (2020) 261}, [\href{https://arxiv.org/abs/1907.10121}{{\tt 1907.10121}}].

\bibitem{ParticleDataGroup:2020ssz}
{\scshape Particle Data Group} collaboration, P.~A. Zyla et~al., \emph{{Review
  of Particle Physics}},
  \href{http://dx.doi.org/10.1093/ptep/ptaa104}{\emph{PTEP} {\bf 2020} (2020)
  083C01}.

\bibitem{Robertson:2016xjh}
A.~Robertson, R.~Massey and V.~Eke, \emph{{What does the Bullet Cluster tell us
  about self-interacting dark matter?}},
  \href{http://dx.doi.org/10.1093/mnras/stw2670}{\emph{Mon. Not. Roy. Astron.
  Soc.} {\bf 465} (2017) 569--587},
  [\href{https://arxiv.org/abs/1605.04307}{{\tt 1605.04307}}].

\bibitem{Kahlhoefer:2015vua}
F.~Kahlhoefer, K.~Schmidt-Hoberg, J.~Kummer and S.~Sarkar, \emph{{On the
  interpretation of dark matter self-interactions in Abell 3827}},
  \href{http://dx.doi.org/10.1093/mnrasl/slv088}{\emph{Mon. Not. Roy. Astron.
  Soc.} {\bf 452} (2015) L54--L58},
  [\href{https://arxiv.org/abs/1504.06576}{{\tt 1504.06576}}].

\bibitem{Brada__2008}
M.~Brada{\v{c}}, S.~W. Allen, T.~Treu, H.~Ebeling, R.~Massey, R.~G. Morris
  et~al., \emph{Revealing the properties of dark matter in the merging cluster
  {MACS} j0025.4-1222}, \href{http://dx.doi.org/10.1086/591246}{\emph{The
  Astrophysical Journal} {\bf 687} (nov, 2008) 959--967}.

\bibitem{Bernal:2019uqr}
N.~Bernal, X.~Chu, S.~Kulkarni and J.~Pradler, \emph{{Self-interacting dark
  matter without prejudice}},
  \href{http://dx.doi.org/10.1103/PhysRevD.101.055044}{\emph{Phys. Rev. D} {\bf
  101} (2020) 055044}, [\href{https://arxiv.org/abs/1912.06681}{{\tt
  1912.06681}}].

\bibitem{Ackermann:2013yva}
{\scshape Fermi-LAT} collaboration, M.~Ackermann et~al., \emph{{Dark Matter
  Constraints from Observations of 25 Milky Way Satellite Galaxies with the
  Fermi Large Area Telescope}},
  \href{http://dx.doi.org/10.1103/PhysRevD.89.042001}{\emph{Phys. Rev. D} {\bf
  89} (2014) 042001}, [\href{https://arxiv.org/abs/1310.0828}{{\tt
  1310.0828}}].

\bibitem{Massey:2015dkw}
R.~Massey et~al., \emph{{The behaviour of dark matter associated with four
  bright cluster galaxies in the 10 kpc core of Abell 3827}},
  \href{http://dx.doi.org/10.1093/mnras/stv467}{\emph{Mon. Not. Roy. Astron.
  Soc.} {\bf 449} (2015) 3393--3406},
  [\href{https://arxiv.org/abs/1504.03388}{{\tt 1504.03388}}].

\bibitem{Modak:2015npa}
K.~P. Modak, \emph{{Constraining Effective Self Interactions of Fermionic Dark
  Matter}},  \href{https://arxiv.org/abs/1509.00874}{{\tt 1509.00874}}.

\bibitem{Jing:2002np}
Y.~P. Jing and Y.~Suto, \emph{{Triaxial modeling of halo density profiles with
  high-resolution N-body simulations}},
  \href{http://dx.doi.org/10.1086/341065}{\emph{Astrophys. J.} {\bf 574} (2002)
  538}, [\href{https://arxiv.org/abs/astro-ph/0202064}{{\tt
  astro-ph/0202064}}].

\bibitem{Simon:2004sr}
J.~D. Simon, A.~D. Bolatto, A.~Leroy, L.~Blitz and E.~L. Gates,
  \emph{{High-resolution measurements of the halos of four dark
  matter-dominated galaxies: Deviations from a universal density profile}},
  \href{http://dx.doi.org/10.1086/427684}{\emph{Astrophys. J.} {\bf 621} (2005)
  757--776}, [\href{https://arxiv.org/abs/astro-ph/0412035}{{\tt
  astro-ph/0412035}}].

\end{thebibliography}\endgroup

\end{document}